\documentclass[english,pra,superscriptaddress,reprint,twocolumn]{revtex4-1}
\usepackage{standalone}
\usepackage{amsmath}
\usepackage{bbold}
\usepackage{nccfoots}
\usepackage{comment}
\usepackage{graphicx}
\graphicspath{{Figures/}}
\usepackage{braket}
\usepackage{color}
\usepackage{nicefrac}
\usepackage{braket}
\usepackage[utf8]{inputenc}
\usepackage{chngcntr}
\usepackage{xr}
\usepackage[normalem]{ulem}

\usepackage[colorlinks=true,
   linkcolor=blue,
   citecolor=blue,
   urlcolor =blue]{hyperref}

\newcommand{\btext}[1]{{\color{black} #1}}

\newcommand{\citeSM}{\cite[{\tiny SM}\kern-0.3em][]{SM}}

\newcommand{\be}{\begin{equation}}
\newcommand{\ee}{\end{equation}}

\expandafter\let\csname equation*\endcsname\relax
\expandafter\let\csname endequation*\endcsname\relax

\begin{document}

\title{Quantum Variational Learning of the Entanglement Hamiltonian}

\author{Christian Kokail${}^{\ast}$}
%\email{Christian.Kokail@uibk.ac.at}
\affiliation{Institute for Quantum Optics and Quantum Information of the Austrian Academy of Sciences, Innsbruck A-6020, Austria}
\affiliation{Center for Quantum Physics, University of Innsbruck, Innsbruck A-6020, Austria}

\author{Bhuvanesh Sundar${}^{\ast}$}
%\email{Bhuvanesh.Sundar@colorado.edu}
\affiliation{Institute for Quantum Optics and Quantum Information of the Austrian Academy of Sciences, Innsbruck A-6020, Austria}
\affiliation{JILA, Department of Physics, University of Colorado, Boulder CO 80309, USA}

\thanks{These authors contributed equally.}

\author{Torsten V. Zache}
%\email{Torsten.Zache@uibk.ac.at}
\affiliation{Institute for Quantum Optics and Quantum Information of the Austrian Academy of Sciences, Innsbruck A-6020, Austria}
\affiliation{Center for Quantum Physics, University of Innsbruck, Innsbruck A-6020, Austria}

\author{Andreas Elben}
%\email{Andreas.Elben@uibk.ac.at}
\affiliation{Institute for Quantum Optics and Quantum Information of the Austrian Academy of Sciences, Innsbruck A-6020, Austria}
\affiliation{Center for Quantum Physics, University of Innsbruck, Innsbruck A-6020, Austria}
\affiliation{Institute for Quantum Information and Matter and Walter Burke Institute for
Theoretical Physics, California Institute of Technology, Pasadena, CA 91125, USA}

\author{Beno\^it Vermersch}
\affiliation{Center for Quantum Physics, University of Innsbruck, Innsbruck A-6020, Austria}
\affiliation{Institute for Quantum Optics and Quantum Information of the Austrian Academy of Sciences, Innsbruck A-6020, Austria}
\affiliation{Univ.~Grenoble Alpes, CNRS, LPMMC, 38000 Grenoble, France}

\author{Marcello Dalmonte}
\affiliation{The Abdus Salam International Center for Theoretical Physics, Strada Costiera 11, 34151 Trieste, Italy}
\affiliation{SISSA, via Bonomea 265, 34136 Trieste, Italy}

\author{Rick van Bijnen}
%\email{rick.van-bijnen@uibk.ac.at}
\affiliation{Institute for Quantum Optics and Quantum Information of the Austrian Academy of Sciences, Innsbruck A-6020, Austria}
\affiliation{Center for Quantum Physics, University of Innsbruck, Innsbruck A-6020, Austria}

\author{Peter Zoller}
%\email{Peter.Zoller@uibk.ac.at}
\affiliation{Institute for Quantum Optics and Quantum Information of the Austrian Academy of Sciences, Innsbruck A-6020, Austria}
\affiliation{Center for Quantum Physics, University of Innsbruck, Innsbruck A-6020, Austria}

\begin{abstract}
Learning the structure of the entanglement Hamiltonian (EH) is central to characterizing quantum many-body states in analog quantum simulation. We describe a protocol where spatial deformations of the many-body Hamiltonian, physically realized on the quantum device, serve as an efficient variational ansatz for a local EH. Optimal variational parameters are determined in a feedback loop, involving quench dynamics with the deformed Hamiltonian as a quantum processing step, and classical optimization. We simulate the protocol for the ground state of Fermi-Hubbard models in quasi-1D geometries, finding excellent agreement of the EH with Bisognano-Wichmann predictions. Subsequent on-device spectroscopy enables a direct measurement of the entanglement spectrum, which we illustrate for a Fermi Hubbard model in a topological phase.
\end{abstract}

\maketitle

\emph{Introduction --} Significant progress has been made in developing quantum simulation hardware \cite{NAP25613, altman2021quantum}. In atomic physics, analog quantum simulators for Bose and Fermi Hubbard models are realized with ultracold atoms in optical lattices \cite{Gross995, sun2020realization, Mazurenko2017, hartke2020doublon, vijayan2020time, holten2021observation, Nichols383, Brown379}, and spin models can be implemented with Rydberg tweezer arrays \cite{scholl2020programmable, ebadi2020quantum, semeghini2021probing} and trapped ions \cite{RevModPhys.93.025001,kokail2019self}. A notable recent development is spatial and temporal control, allowing addressing of single lattice sites, and single-shot single-site read out of atoms~\cite{NAP25613,altman2021quantum}, e.g.~as spin and density resolved measurements with a quantum gas microscope \cite{gross2020quantum, Lukin256}. The generic many-body Hamiltonian realized in analog quantum simulators has a (quasi-) local structure, $\hat H(\boldsymbol{g})=\sum_{i}g_{i}\hat h_{i}$, where the $\hat h_{i}$ act non-trivially on spatially contiguous sites $i$ as few-body operators. Achieving local control thus implies tunability of the spatial couplings $g_{i}$. Analog quantum simulators are, therefore, capable of not only realizing homogeneous, i.e.~`in-bulk' translationally invariant Hamiltonians $\hat H=\sum_{i}\hat h_{i}$, but a whole family of spatially \emph{deformed Hamiltonians} $\hat H(\boldsymbol{g})$ with a spatiotemporally programmable pattern $\boldsymbol{g}\!\equiv\!\{g_{i}\}$. 
This programmability provides us with opportunities to design specific classes of quantum protocols, running on the quantum simulator, to achieve tasks of interest in quantum many-body physics. 
\begin{figure}[!ht]
\includegraphics[width=1.0\columnwidth]{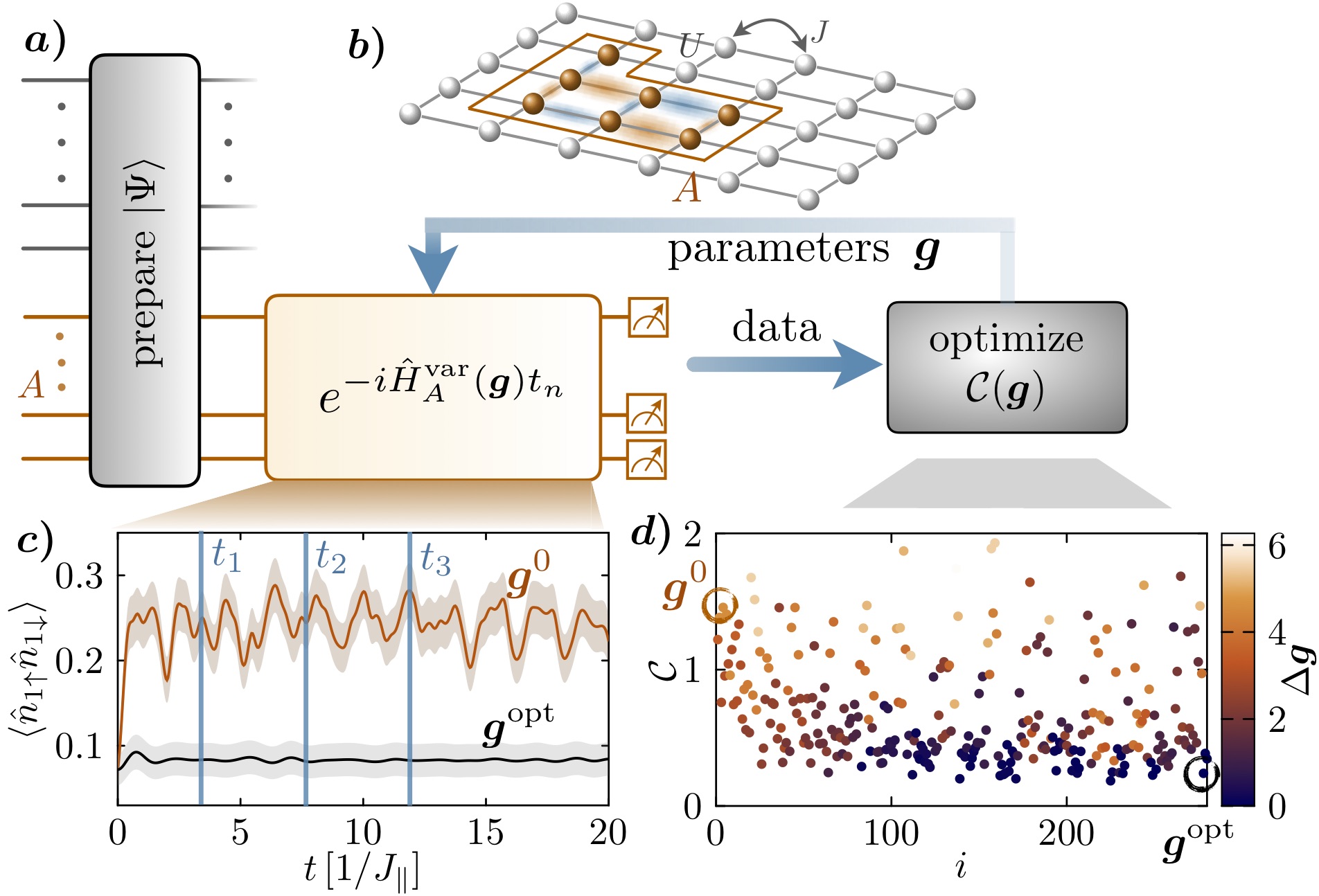}
\caption{\textit{Quantum variational learning (QVL) of EH} a) subsystem $A$ is time-evolved with the deformed Hamiltonian $\hat{H}_A^\text{var}(\boldsymbol{g})$, while measuring observables $\braket{\hat{\mathcal{O}}_A}_t$  at time instances $\{ t_n \}$. A classical computer optimizes a cost function $\mathcal{C}(\boldsymbol{g})$ in a feedback loop, minimizing the time variation of observables. b) Fermi Hubbard  Model, and subsystem $A$.  c) Time variation of observables indicating convergence in the feedback loop from the initial $\boldsymbol{g}^0$ to final $\boldsymbol{g}^{\rm opt}$ parameters, d) corresponding cost function vs.~iteration number of the optimizer. The colormap visualizes the distance to the final parameter vector $\Delta \boldsymbol{g} = |\boldsymbol{g}^i - \boldsymbol{g}^\text{opt}|$. Data plotted in c) and d) were obtained in simulated runs for a 2-leg Fermi-Hubbard ladder [see text and Fig.~2a),b)], monitoring double site occupancy by means of a quantum gas microscope.
}
\label{fig:protocol}
\end{figure}

Below we describe a protocol based on a hybrid classical-quantum algorithm \cite{cerezo2020variational} to learn the entanglement Hamiltonian (EH) of a subsystem $A$ of a quantum many-body state [see Eq.~\ref{eq:reduced_density_operator}]. In the protocol, a \textit{deformed Hamiltonian} $\hat H_A^\text{var}(\boldsymbol{g})$ plays the role of an \textit{ansatz for the EH}, where $\boldsymbol{g}$ represents a small set of variational parameters scaling polynomially with the system size. These are determined efficiently in a quantum feedback loop from monitoring the time evolution of certain local, experimentally accessible observables evolving under $\hat H_A^\text{var}(\boldsymbol{g})$. 
As outlined in Fig.~\ref{fig:protocol}a), our protocol differs from classical learning (CL) methods \cite{bairey2019learning, qi2019determining, li2020hamiltonian, evans2019scalable} by implementing a quantum processing step through time evolution with the deformed Hamiltonian, acting \textit{in situ} on the quantum state stored in quantum memory of the quantum simulator.  A unique feature of the present setting is that the learned EH is also available as a physical Hamiltonian on the quantum device for further experimental studies,  
\btext{
such as, e.g., determining the entanglement spectrum (ES) through spectroscopy. This is in contrast to tomography-based methods ~\cite{kokail2020entanglement, choo2018measurement}, where the ES is obtained by diagonalizing the (learned) EH on a classical computer.
}

% {\color{blue}We emphasize that our focus is learning the EH and finding the ES in analog quantum simulators, using only resources native to the device, in contrast to circuit-based protocols~\cite{peruzzo2014variational,larose2019variational, johri2017entanglement, subacsi2019entanglement, cerezo2020variational2} for finding the ES, and quantum algorithms for matrix diagonalization~\cite{lloyd2014quantum, bravo2020quantum}. Our protocol scales to regimes beyond what is possible using protocols relying on classical post-processing~\cite{kokail2020entanglement, choo2018measurement}, i.e., diagonalizing the (learned) EH on a classical computer. 
% }

 \btext{
 %While there exist general digital algorithms designed for universal quantum computers ~\cite{peruzzo2014variational,larose2019variational, johri2017entanglement, subacsi2019entanglement, cerezo2020variational2, lloyd2014quantum, bravo2020quantum}, these require a significant overhead for realizing fermionic systems.  As we illustrate below, our protocol makes use of the resources naturally available on analog quantum simulators, providing an efficient approach, e.g., for studying entanglement in Fermi-Hubbard models. 
% Our protocol scales to regimes beyond what is possible using protocols relying on classical post-processing~\cite{kokail2020entanglement, choo2018measurement}, i.e., diagonalizing the (learned) EH on a classical computer. 
We emphasize that by devising variational quantum algorithms in the framework of analog simulation we build on existing, scalable and high-fidelity quantum hardware, capable of realizing physically motivated variational ansätze for the EH. As illustrated below, this hardware efficiency includes the ability to represent fermions in Hubbard models naturally as fermionic atoms and associated fermionic quantum operations. While digital algorithms~\cite{peruzzo2014variational,larose2019variational, johri2017entanglement, subacsi2019entanglement, cerezo2020variational2, lloyd2014quantum, bravo2020quantum} offer in principle a broader scope of applicability, they come in general with the significant hardware requirement of a freely programmable quantum computer and involve a technical overhead for realising fermionic models.

%We contrast the analog approach to digital postprocessing, based on running quantum algorithms such as .... While this digital approach offers in principle a broader scope of applicability, it come in general with the significant quantum hardware requirement of a freely programmable quantum computer.
}

\emph{Ansatz for EH as deformed Hamiltonian --} The entanglement Hamiltonian (EH) $\tilde{H}_A$ and the collection of its eigenvalues $\{\xi_{\alpha}\}$, the entanglement spectrum (ES), are central to our understanding of complex quantum states as they completely characterize all correlations in a subsystem $A$. Given a many-body state $\hat \rho$, they are related to the reduced density matrix on $A$ via 
\begin{equation}\label{eq:reduced_density_operator}
  \hat \rho_{A} \equiv \mathrm{Tr}_{\neg A}[ \hat \rho] \equiv \exp (-\tilde H_A) =\sum_{\alpha}e^{-\xi_{\alpha}}\ket{\Phi_{A}^{\alpha}}\bra{\Phi_{A}^{\alpha}} .
\end{equation} 
The ES can distinguish different quantum phases, e.g. its low-lying part reflects the structure of the conformal field theory (CFT) describing edge excitations in a topological phase \cite{regnault2015entanglement, haag2012local}. Moreover, the EH plays a key role in the holographic approach to geometry emerging from entanglement \cite{Casini:2011aa}.

In many physically relevant cases, $\tilde{H}_A$ is a \emph{deformation} of the system Hamiltonian $\hat H$.
A seminal example is provided by the Bisognano-Wichmann (BW) theorem of local quantum field theory (QFT)~\cite{bisognano1975duality}. It states that the EH for the ground state of a relativistic QFT and a subsystem $A$ defined by $x_1 > 0$ is given by \mbox{$\tilde H_A =  \int_{{\mathbf x} \in A} d{\mathbf x} \, \beta({\mathbf x}) {\cal \hat{H}} ({\mathbf x}) \!+\! c$. Here ${\cal \hat{H}} ({\mathbf x})$} is the energy density of $\hat H$, $c$ is a normalization constant and the EH is parametrized by a local ``inverse temperature''  $\beta({\mathbf x})=2\pi x_1$, taking the form of a linear ramp. We emphasize that the BW theorem holds in arbitrary spatial dimensions \footnote{For a generalization within CFT to finite subsystems of radius $R$, see Ref.~\cite{hislop1982,Casini:2011aa,Cardy_2016} who proved that the deformation takes a parabolic shape, $\beta({\mathbf x})= 2\pi (R^2 - \mathbf{x}^2)/(2R)$, or similarly with the chord length.}. Remarkably, BW-like deformations also provide excellent approximations for the EH of the ground state in a variety of lattice models~\cite{pourjafarabadi2021entanglement, eisler2020entanglement,giudici2018entanglement,Toldin:2018aa}.
Based on this observation, Ref.~\cite{dalmonte2018quantum} proposed that, \emph{assuming} the validity of a lattice version of the BW theorem, the BW-deformed Hamiltonian can be physically realized and probed in quantum simulation experiments.
In contrast, our hybrid classical-quantum learning algorithm explicitly \textit{finds} the optimal variational approximation for the EH among a class of deformed system Hamiltonians.

\emph{Protocol --} The key ingredient of the algorithm is the capability of the quantum simulator to realize unitary evolution under deformed Hamiltonians \mbox{$\hat H^\textrm{var}_A(\boldsymbol{g}) = \sum_{j\subset A} g_j \hat h_j$}, acting for some time $t$ on a subsystem of interest $A$. As illustrated in Fig.~\ref{fig:protocol}a, we first prepare a desired quantum state, then evolve the subsystem according to $\hat H^\textrm{var}_A(\boldsymbol{g})$, and monitor the evolution of \btext{local} observables $\mathcal{\hat O}_A$ in the subsystem,
\begin{equation}\label{eq:constraint}
    \langle \mathcal{\hat O}_A \rangle_t \equiv \text{Tr}_A \left[\mathcal{\hat O}_A e^{-i \hat H^\textrm{var}_A(\boldsymbol{g}) t} \hat \rho_A e^{ i \hat H^\textrm{var}_A(\boldsymbol{g}) t} \right].
\end{equation}

The classical-quantum feedback loop consists in finding an optimal set
$\boldsymbol{g}^\text{opt}$ by minimizing the time variation of the observables, i.e. we wish to enforce $\langle \mathcal{\hat O}_A \rangle_t = \textrm{const}$. In practice, we achieve this by minimizing a cost function \mbox{$\mathcal{C}(\boldsymbol{g})= \sum_{t \in T,\hat{\mathcal{O}}_A \in O} \left(\langle \hat{\mathcal{O}}_A \rangle_t - \langle \mathcal{\hat O}_A \rangle_0\right)^2$}, \btext{where $T=\{t_i\}$ denotes a set of observation times. The precise choice of observables $\hat{\mathcal{O}}_A$ is not critical for our protocol, as we expect the quantum dynamics to scramble them into complex many-body operators as long as $\left[\hat{ \mathcal{O}}_A,\hat{H}^\textrm{var}_A(\boldsymbol{g})\right] \ne 0$. }
%Monitoring sufficiently many observables $\{\mathcal{\hat O}_A\}$ providing constraints, the algorithm finds the optimal variational approximation of the EH.
\btext{Thus, monitoring a small number of local observables at different observation times $\{ t_i \}$ provides a sufficient number of constraints for the algorithm to find an optimal variational approximation to the EH.}
This is efficient in view of the quasi-local ansatz with a small set of variational parameters, and we refer to Appendix \ref{app:alg} for a detailed technical discussion, including the choice of observables and the role of conservation laws. 

We note that the EH is obtained from Eq.~(\ref{eq:constraint}) only up to a scale factor and an overall shift, $ \tilde{H}^\textrm{var}_A = \beta \hat H(\boldsymbol{g}^\text{opt}) + c$, i.e.~the ES is determined as universal ratios, \mbox{$\kappa_\alpha=(\xi_\alpha-\xi_{\alpha_0})/ (\xi_{\alpha_1}-\xi_{\alpha_0})$}. As discussed in Appendix section \ref{app:scale}, these scale factors can be determined in additional steps.

\emph{Learning the EH of ground states of the Fermi-Hubbard model --}  We now demonstrate the quantum EH learning protocol for the Fermi-Hubbard model (FHM). The FHM is a paradigmatic model in condensed matter physics for a strongly interacting quantum many-body system, and in two spatial dimensions (2D) is central to studies of high-temperature superconductivity. The FHM is described by the Hamiltonian
\begin{align}\label{eq:FHM}
\begin{split}
\hat{H}_{\text{FHM}} = &- J \sum_{\braket{jk}, \sigma}  \left(\hat{c}^{\dagger}_{j \sigma} \hat{c}_{k \sigma} + \text{H.c.} \right) \\ & + U \sum_{j} \hat{n}_{j\uparrow}  \hat{n}_{j\downarrow}  - \mu \sum_{j\sigma}  \hat{n}_{j \sigma}  \;,
\end{split}
\end{align}
with $\hat{c}_{j\sigma}$ ($\hat{c}^\dagger_{j\sigma}$) destruction (creation) operators for fermions on lattice site $j$ with spin $\sigma=\{\uparrow,\downarrow\}$. The first term describes hopping of particles with tunneling strength $J$ between neighboring sites $\braket{j k}$, the second term represents an on-site interaction with strength $U$ with densities $\hat{n}_{j \sigma} = \hat{c}^\dagger_{j\sigma} \hat{c}_{j\sigma}$, and the last term involves chemical potentials $\mu_{\sigma}$. The FHM is  realized in state-of-the-art quantum simulators employing fermionic atoms trapped in optical lattices \cite{Mazurenko2017, hartke2020doublon, vijayan2020time, holten2021observation}. 

\begin{figure}[t]
\includegraphics[width=1.0\columnwidth]{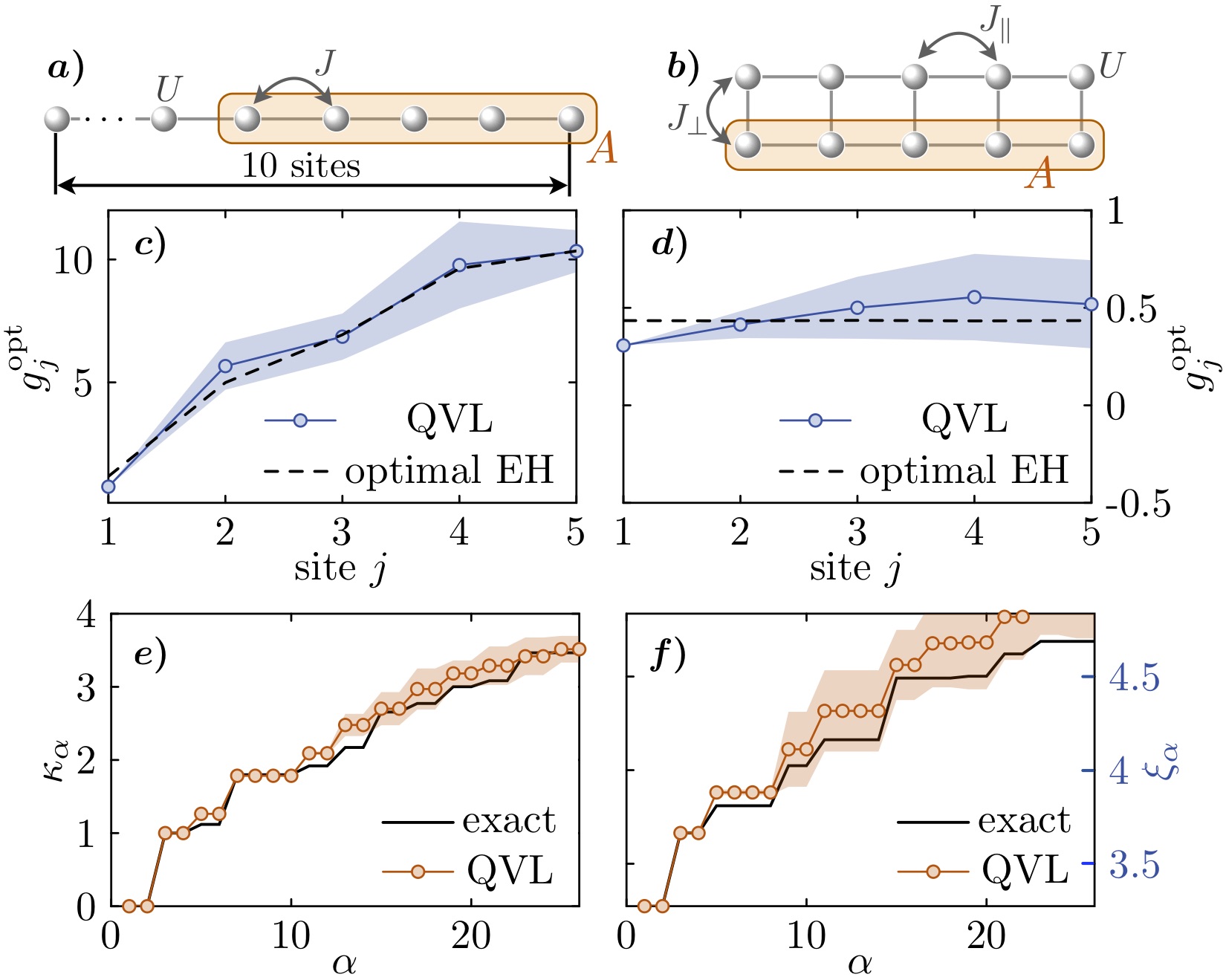}
\caption{\emph{Quantum variational learning (QVL) of the EH for different geometries of a Hubbard model.} Left column: results for a 5-site subsystem on the right boundary of a 10-site Hubbard chain with $U/J=1$ and $\mu/J = -0.5$. Right column: results for a single chain in a 2-leg ladder with $U/J_\parallel=8$, $J_\bot/J_\parallel = 2$ and $\mu/J = 0$. a), b) lattice geometries with highlighted subsystems. c), d) variational parameters $g_j^\text{opt}$ obtained from optimization with $6\times 10^4$ experimental runs (QVL), fixing $g_{j=1}$. The optimal parameters are rescaled (corresponding fidelities shown in Fig.~\ref{fig:Fid}) for comparison with the EH parameters obtained by numerically optimizing the relative entropy (black solid lines, see also Appendix \ref{app:relent}). e), f) Universal ratios $\kappa_\alpha$ calculated by diagonalizing the variationally obtained EH $\hat{H}^\text{var}_A(\boldsymbol{g}^\text{opt})$ in comparison to exact eigenvalues of the reduced density matrix $\rho_A$.
All simulations are performed for the ground state in the zero magnetization sector. Error bands are computed by repeating the entire optimization run 10 times and computing the standard error.}
%\hat{h}_j = -J \!\!\!\!\! \sum_{\sigma, \ell = \{ \pm 1 \}} \!\!\! ( \hat{c}_{j+ \ell\, \sigma}^{\dagger} \hat{c}_{j\, \sigma} + \text{H.c.} ) + U \hat{n}_{j \uparrow} \hat{n}_{j \downarrow} - \mu \sum_{\sigma} \hat{n}_{j \sigma}}
\label{fig:FermiHubbard}
\end{figure}

We illustrate quantum variational learning (QVL) of the EH structure for  the FHM with two examples (see Fig.~\ref{fig:FermiHubbard}). The first example considers a 1D chain with subsystem $A$ on the right boundary [Fig.~\ref{fig:FermiHubbard}a)]. The second example is a two-leg ladder, which is cut horizontally defining $A$ as the lower leg [Fig.~\ref{fig:FermiHubbard}b)]. For the two-leg ladder, we consider a slight modification to the FHM with anisotropic hopping $J \to J_\parallel, J_\perp$ between horizontal and vertical links, respectively. In both FHM examples, we assume the total system is in its ground state with half filling, and in the zero magnetization sector. As an ansatz for the deformed Hamiltonian to be learned for these examples, we choose $\hat{H}^{\rm var}(\boldsymbol{g}) = \sum_{j\in A} g_j \hat{h}_j$, defined on a subsystem $A$, with quasi-local operators centered on lattice site $j$:
\begin{align} \label{eq:hj}
%\hat{h}_j = -\hspace{-0.3cm}\sum_{\substack{k \in \braket{jk}\cap A \\ \sigma}} 
\hat{h}_j = -\hspace{-0.3cm}\sum_{k \in \braket{jk}\cap A}\sum_\sigma
\frac{J}{2}(\hat{c}^{\dagger}_{j \sigma} \hat{c}_{k \sigma} + \text{H.c.}) +  U \hat{n}_{j\uparrow}  \hat{n}_{j\downarrow}  - \mu \sum_{\sigma}  \hat{n}_{j \sigma},
\end{align}
where for the horizontally cut ladder, $J = J_\parallel$.
We note that for the full system, $\hat{H}_{\text {FHM}} \equiv \sum_j \hat{h}_j$, and that the ansatz $\hat{H}^{\rm var}(\boldsymbol{g}) = \sum_{j\in A} g_j \hat{h}_j$ can thus be viewed as a discretized lattice version of the BW deformation, as originally defined in the continuum. Realizing such a deformed Hamiltonian in the laboratory requires local control over the Hamiltonian parameters $J, U$, and $\mu_\sigma$, which can be achieved e.g. by using digital mirror devices to shape optical potentials \cite{qiu2020precise}, and through Raman induced laser couplings \cite{goldman2016topological}. Alternatively, time evolution with a deformed Hamiltonian can also be naturally implemented as digital quantum simulation, achieved with spatially homogeneous Hamiltonians acting for short times on properly chosen subregions of $A$ (see Appendix \ref{app:trotter}).

%Alternatively, time evolution with a deformed Hamiltonian can also be naturally implemented as digital quantum simulation, i.e.~decomposed into a sequence of Trotter steps. As shown in SM, each Trotter step $i$ is generated by a FHM (\ref{eq:FHM}) with spatially homogeneous $J,U$ restricted to a properly chosen subregion $H \mid_{A_i \subset A}$ for a time $\Delta t_i$. 
%%This digital approach, as sequence of local Hubbard quenches representing Fermi many-particle quantum gates, has a natural physical realization in atomic setups. 
%Finally, the protocol requires the identification of a set of experimentally accessible observables, providing sufficient constraints to learn the EH. As shown below, for the atomic Hubbard model this includes spin- and site-resolved local densities as measured with a quantum gas microscope, or measurement of local atomic currents. 

We numerically simulate the full protocol of determining the EH (Fig.~1), including quantum projective measurements and variational optimization with an adaptive DIRECT algorithm, as used in Ref.\ \cite{kokail2019self}, constraining the total number of experimental runs to $6 \times 10^4$. As observables to be monitored, we choose the double occupancy on lattice sites for the first example \footnote{\label{footnote:g_constraint}We note that here the variational search is constrained to $g_j > 0$ for all $j$. In the case that $g_j < 0$ are allowed, a solution with $g_j = -g_{j+1}$ does exits which freezes the particles on the individual lattice sites. To exclude this solution, additional observables like nearest-neighbour tunneling amplitudes are required.}, and for the second example local tunneling elements $\mathcal{J}^{\sigma}_{j, j+1} =   \hat{c}_{j \sigma}^{\dagger} \hat{c}_{j+1,\sigma} + \text{H.c.}$, which can be accessed by inducing super-exchange oscillations accompanied by site-resolved measurements in a quantum gas microscope \cite{PhysRevLett.117.170405, PhysRevA.89.061601}.

For the 1D Hubbard chain, Fig.~\ref{fig:FermiHubbard}c) shows the optimized parameters $\boldsymbol{g}^\text{opt}$, consistent with the BW expectation of an approximately linear ramp, but bending over to a parabolic shape due to boundary effects. For the two-leg FHM with horizontal cut, Fig.~\ref{fig:FermiHubbard} d) shows the learned deformation as approximately flat, again in agreement with a minimal version of BW. We can understand this result perturbatively in the limit $U \gg J_{\perp/||}$ for $J_\perp \gg J_{||}$. In this case, following \cite{PhysRevB.85.054403}, the EH is proportional to the system Hamiltonian restricted to a single leg of the ladder. 

Having learned the operator structure of the EH $\hat H^{\rm var}(\boldsymbol{g}^{\rm opt})=\sum_j g^{\rm opt}_j \hat h _j$, and having a realization of the EH available as physical Hamiltonian on the quantum device, we can proceed to extract entanglement properties encoded in the EH with both classical or quantum (on device) postprocessing. Below we focus on the entanglement spectrum, which is obtained either by diagonalizing the EH classically, or via `on device' spectroscopy, which potentially scales to regimes beyond classical postprocessing.  

\begin{figure}[t]
\includegraphics[width=1.0\columnwidth]{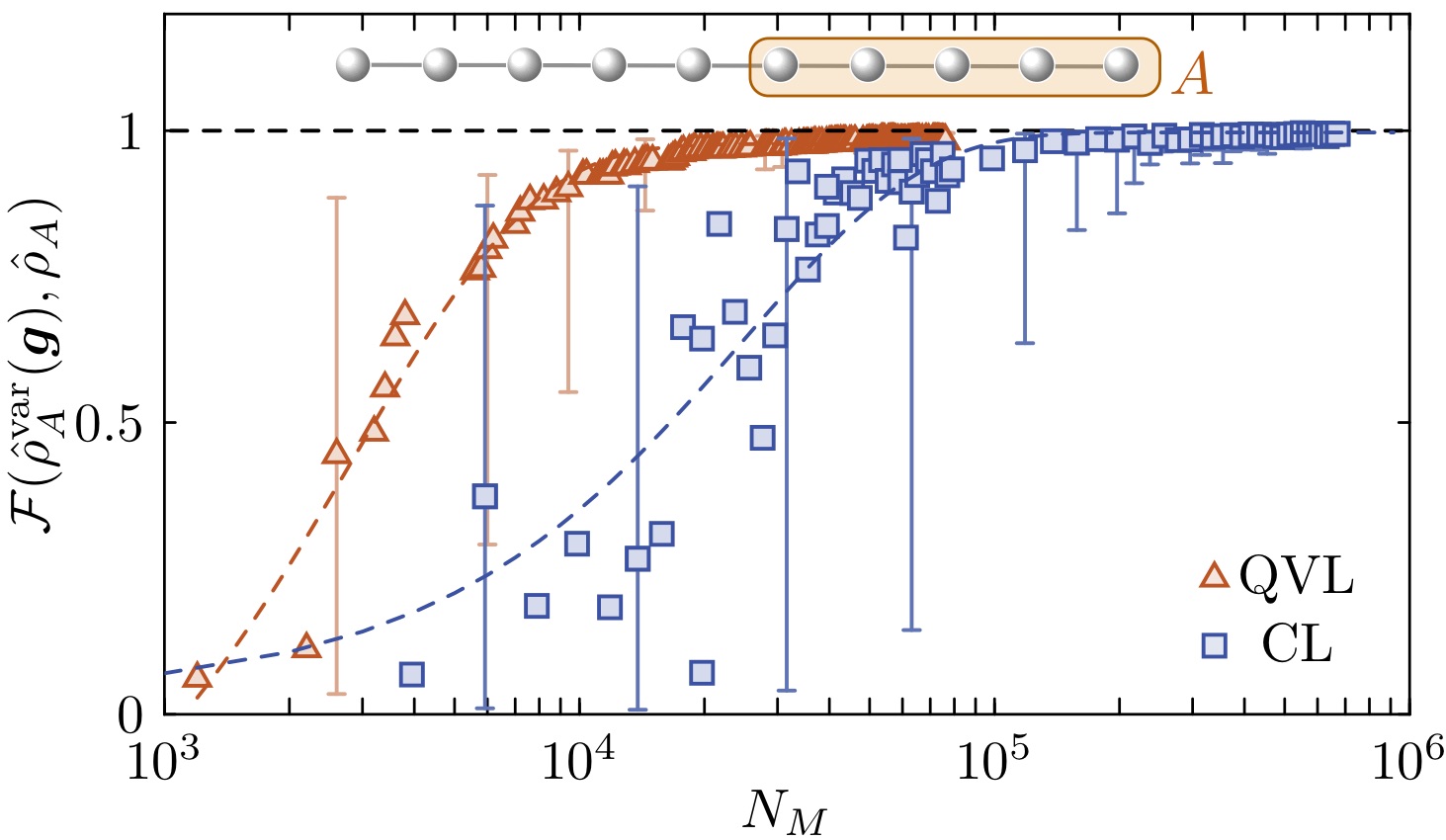}
\caption{\textit{Error assessment vs.~number of experimental runs $N_M$:}  Maximally achievable Uhlmann fidelity with respect to the exact density matrix $\hat{\rho}_A$ as a function of the total number of measurements for a half-partition of a 10-site Hubbard chain ($N_A = 5$) for $U/J=1$ as shown in the inset.
For comparison, the blue data points represent a learning protocol based on classical post-processing of measurement data \cite{bairey2019learning}. The data show the median of the fidelity when the experiment is repeated 100 times. We plot a selection of representative error bars which indicate $2\sigma$ confidence intervals.}
\label{fig:Fid}
\end{figure}

{\em Classical postprocessing of the EH --}  Fig.~\ref{fig:FermiHubbard}e,f) shows universal ratios $\kappa_\alpha$ obtained by diagonalizing the learned EH. The results compare favorably to the exact values within 2$\sigma$ error bands. 
To further quantify the performance of the EH reconstruction, we compare in Fig.~\ref{fig:Fid}b) the Uhlmann fidelity $\mathcal{F}(\hat{\rho}^\text{var}_A(\boldsymbol{g}), \hat{\rho}_A)$ \footnote{To be precise, we consider a family of reconstructed states $\hat{\rho}^\text{var}_A(\boldsymbol{g}, \beta) = Z^{-1}_A(\beta) \, e^{-\beta \hat{H}^\text{var}_A(\boldsymbol{g})}$ with $Z_A(\beta) = \text{Tr} \left[e^{-\beta \hat{H}^\text{var}_A(\boldsymbol{g})}\right]$, and define the fidelity as $\mathcal{F}(\hat{\rho}^\text{var}_A(\boldsymbol{g}), \hat{\rho}_A) = \max_\beta \left[ \text{Tr} (\sqrt{\sqrt{\hat{\rho}_A} \hat{\rho}^{\text{var}}_A(\boldsymbol{g},\beta) \sqrt{\hat{\rho}_A}}) \right]^2$.} of the reconstructed state $\hat{\rho}^\text{var}_A(\boldsymbol{g})$
with respect to the exact density matrix $\hat{\rho}_A$ as a function of the total number of experimental runs.
%and for different learning protocols and EH parametrizations~\footnote{To be precise, we consider a family of reconstructed states $\hat{\rho}^\text{var}_A(\boldsymbol{g}, \beta) = Z^{-1}_A(\beta) \, e^{-\beta \hat{H}^\text{var}_A(\boldsymbol{g})}$ with $Z_A(\beta) = \text{Tr} \left[e^{-\beta \hat{H}^\text{var}_A(\boldsymbol{g})}\right]$, and define the fidelity as $\mathcal{F}(\hat{\rho}^\text{var}_A(\boldsymbol{g}), \hat{\rho}_A) = \max_\beta \left[ \text{Tr} (\sqrt{\sqrt{\hat{\rho}_A} \hat{\rho}^{\text{var}}_A(\boldsymbol{g},\beta) \sqrt{\hat{\rho}_A}}) \right]^2$.}.
The analysis is performed for a 5-site subsystem on the right boundary of a 10-site Hubbard chain as depicted in Fig.~\ref{fig:Fid}. We present results for a parametrization of the form $\hat{H}_A^\text{var}(\boldsymbol{g}) = \sum_{j \in A} g_j \hat{h}_j$, with operators $\hat{h}_j$ as defined in Eq.~(\ref{eq:hj}), which reaches fidelities close to 1 with a remarkably small number $N_M \sim \mathcal{O}(10^4)$ of experimental runs. %As shown in the inset, we initialize the variational search with a negative slope for the parameters, opposite to the expected optimal ones. 
In our numerical experiments, we initialize each variational search with a random parameter vector $\boldsymbol{g}^{0}$.
%To further illustrate the performance of our protocol, the green curve in Fig.~\ref{fig:betaFid}b) shows results corresponding to a second parametrization of the form $\tilde{H}_A(\boldsymbol{g}) = \sum_j ( g_j^{(1)} \hat{h}_j^{(1)} + g_j^{(2)} \hat{h}_j^{(2)})$, where $\hat{h}_j^{(1)}$ represents the operators proportional to $J_{\hat{e}}$ in Eq.~(\ref{eq:local_operators_FHM}) and $\hat{h}_j^{(2)} = \hat{h}_j - \hat{h}_j^{(1)}$. Here the links of the lattice are deformed independently of the components acting on the lattice sites, resulting in 9 variational parameters and thus a higher total measurement budget.

%Finally, 
The blue curve in Fig.~\ref{fig:Fid}b) shows the behavior of the Uhlmann fidelity for a {\em classical protocol} to learn the EH as developed by \cite{bairey2019learning} for system Hamiltonians, which we adapt here to EHs. This approach is based on measuring local observables $\hat{\mathcal{O}}_A$ which involve next- and next-next-nearest neighbor atomic currents (see Appendix \ref{app:comp}). This is in contrast to QVL, where measurement of nearest-neighbour currents and local densities is sufficient. Fig.~\ref{fig:Fid} shows results, where we estimate the scaling with a finite number of runs $N_M$ by adding independent Gaussian noise to the observables $\mathcal{O}_A$, with zero mean, and variance $\epsilon^2 = \text{Var} (\mathcal{O}_A)/N_M$ (see Appendix \ref{app:comp}). 
%For this setup, the classical approach requires a drastically increased measurement budget. 
While QVL is bound to the restriction of implementing deformed Hamiltonians on the quantum device, convergence is achieved significantly earlier compared to CL.
We note that for CL the number of experimental runs may be reduced by a factor $\sim \! N_A$ by grouping operators $\mathcal{O}_A$ into commuting sets which can be measured simultaneously.

\begin{figure}[t]
\includegraphics[width=1.0\columnwidth] {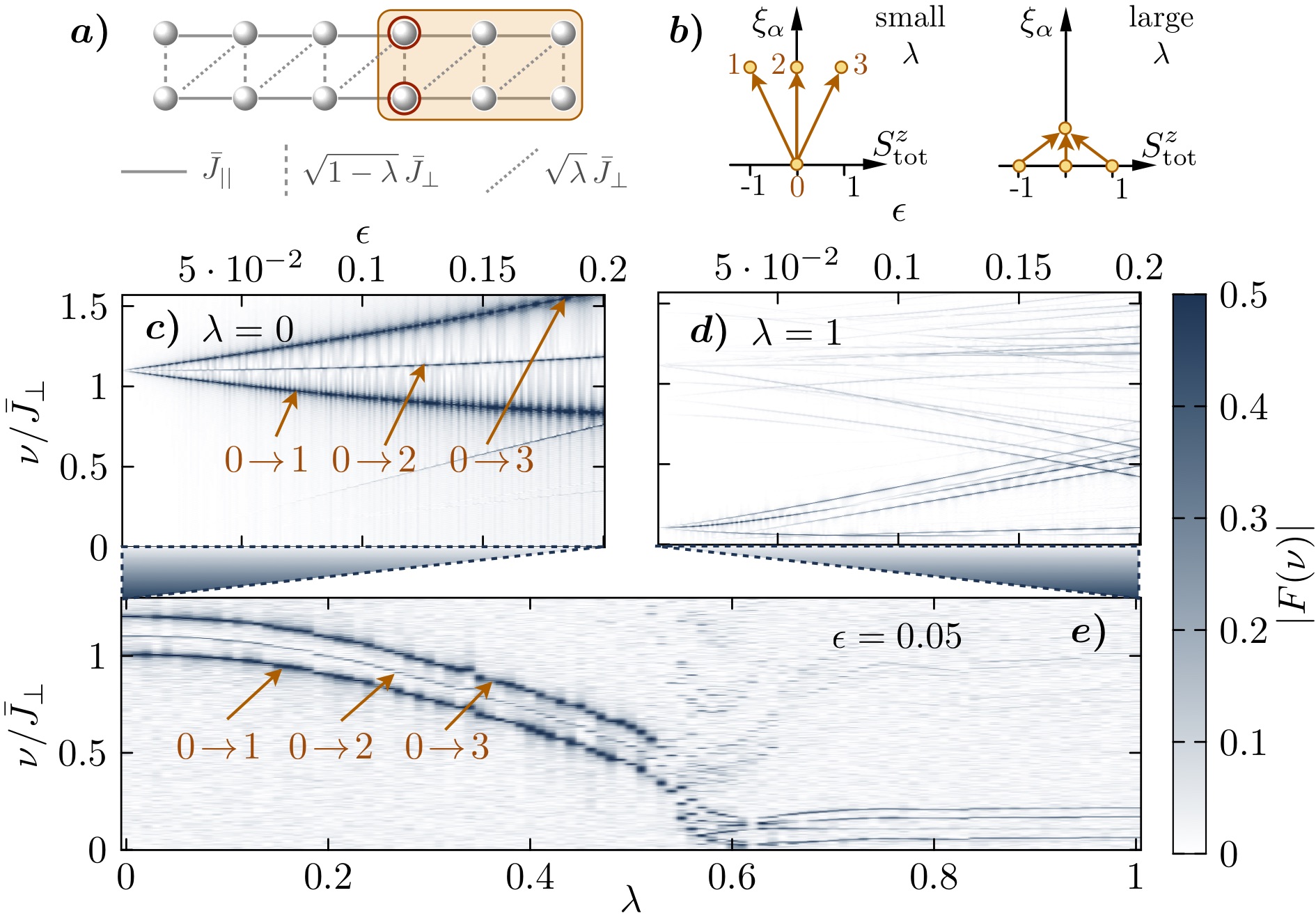}
\caption{ {\em Entanglement spectroscopy of the Heisenberg model on a ladder.} a)  Schematic of the setup, with subsystem $A$ the right half of a $12$-site ladder. %The sites on the top left and bottom right corners couple to their neighbors with strength $\lambda \bar{J}$.
\btext{ We fix the ratio of horizontal to vertical couplings as $J_{||}/J_\perp = 0.25$ and tune the relative strength of diagonal and vertical couplings with $\lambda \in [0,1]$.}
b) Level scheme of the low-lying part of the entanglement spectra, indicating the dominant transitions\btext{illustrated here for $\epsilon = 0$}. 
c) Measured entanglement spectrum (see main text), for small values $\epsilon$ of the perturbation, in the %topological
\btext{ trivial} phase ($\lambda = 0)$. d) Same as c) at $\lambda=1$. %The next ground-excited transitions appear at $\nu/\bar{J} \sim 2.2$
e) Measured entanglement spectra versus $\lambda$, at $\epsilon = 0.05\bar{J}_\perp$. %The lowest spectral line shift to $\nu=0$ at $\delta=0$, indicating a topological phase there -- the symmetry-protected Haldane phase.
All spectra are computed assuming $5000$ measurements per observable at each time, measured up to $t = 1000/\bar{J}_\perp$. }
\label{fig: ES}
\end{figure}

\emph{On-device entanglement spectroscopy -- } 
% As a result of our variational protocol, we learn how to realize the EH directly on the quantum simulator. This provides us with the unique opportunity to probe the EH using quantum post-processing on the same device.
The realization of the EH as a physical Hamiltonian on the quantum device allows measurement of the ES via spectroscopy \cite{dalmonte2018quantum} (see also \cite{pichler2016measurement}).
Below we illustrate such a quantum post-processing step and simulate entanglement spectroscopy. To this end, we evolve the reduced system $\hat{\rho}_A$ once again, but now with a perturbation added to the EH, \mbox{$\hat{H}_A^\text{var}( \boldsymbol{g}^\text{opt} ) + \epsilon \hat{H}'$}. For an appropriately chosen weak perturbation $\epsilon\hat{H}'$, the subsystem's response exhibits a quantum beat pattern with frequencies $\propto \left(\xi_\alpha - \xi_\beta\right)$ that can be extracted by extrapolating $\epsilon \rightarrow 0$. 

%The entanglement spectrum can be obtained by measuring the dynamic response of the system to a perturbation, i.e., evolving the system with $\hat{H}( {\bf g}^\text{opt} ) + \epsilon \hat{H'}$ where $\epsilon \hat{H'}$ is a perturbation, and making measurements during the resulting dynamics. The measurements show a beat pattern with frequencies $\xi_\alpha - \xi_\beta$ in the limit of small $\epsilon$, where $\xi_{\alpha,\beta}$ are eigenvalues of the learned EH (see SM for details).

For simplicity, we consider the FHM on a ladder geometry [see Fig.~\ref{fig: ES}a)] in the limit of large on-site interaction $U\gg J$, where it reduces to the Heisenberg model described by the Hamiltonian
\begin{align}
\hat{H} = \sum_{\langle i j \rangle} \hat{h}_{ij} \;, &&
\hat{h}_{ij} = \bar{J}\sum_{a = x,y,z} \hat{\sigma}^a_i \hat{\sigma}^a_j  \;,
\end{align}
with the sum running over neighboring sites on the ladder and we abbreviated $\bar{J}=J^2/(2U)$. We first apply our protocol to find an optimal EH, $\hat{H}_A^{\rm var}(\boldsymbol{g}^\text{opt} ) = \sum_{\langle ij \rangle_A} g^\text{opt}_{ij}\hat{h}_{ij}$, for the subsystem $A$ indicated in Fig.~\ref{fig: ES}a). The expected level structure of the corresponding ES is shown in Fig.~\ref{fig: ES}b). 
In order to induce transitions that resolve the degeneracy of the low-lying levels, we then evolve with the learned EH perturbed by a local magnetic field, $\hat{H}' = \sum_{i_0=1,2}\vec{B}_{i_0} \cdot \hat{\vec{\sigma}}_{i_0}$, supported on the two sites $i_0=1,2$ at the edge of the entanglement cut indicated by the red circles in Fig.~\ref{fig: ES}a), with $\vec{B}_1 =(1,0,1)$ and $\vec{B}_2 = (1,0,-1/2)$.

We probe the response of the system by measuring $f(t) = \braket{ \sum_{j \in A} (-1)^{j_x+j_y} \hat{\sigma}^z_j(t)}$ and plot the corresponding discrete cosine transformed spectrum $F(\nu)$ in Fig.~\ref{fig: ES}c). The dominant lines correspond to transitions between the ground state and the first three excited states of $\tilde{H}_A$. Beating between excited states is significantly weaker due to the thermal occupation in the state $\hat{\rho}_A = \exp\left(-\tilde{H}_A \right)$. Our results clearly demonstrate that the values $\xi_{1,2,3} - \xi_0$ can be obtained by extrapolating the peak positions to $\epsilon=0$. Importantly, the Zeeman-type splitting provides a clear resolution of the three-fold degeneracy. \btext{ In an experiment, the ability to resolve this splitting will be limited by the %maximal
coherence time of the device.}

Measuring the ES and resolving its degeneracies constitutes a powerful tool to distinguish different quantum phases and identify topological order.
Motivated by a recent experiment~\cite{sompet2021realising}, we demonstrate this possibility in a generalized model, %where two lattice sites on the top left and bottom right corners are coupled to their neighbors with a modified strength $\lambda \bar{J}$ as indicated by the weak links in Fig.~\ref{fig: ES}a).
\btext{ where we decrease the inter-leg couplings $\sqrt{1-\lambda}  \, \bar{J}_\perp$ while increasing new diagonal terms with strength $\lambda \, \bar{J}_\perp$, as indicated by the dashed and dotted links in Fig.~\ref{fig: ES}a). This situation can be realized experimentally by displacing the two legs along the longitudinal direction, thereby smoothly interpolating between the previous analysis at $\lambda=0$ and a Haldane phase at large $\lambda$.}
%While the previous analysis corresponds to $\lambda = 1$, the case of  $\lambda=0$ realizes the Haldane phase.
According to the Li-Haldane conjecture~\cite{li2008entanglement}, this topological phase can be directly detected with the ES by counting the degeneracy of the ground state of $\tilde{H}_A$. \btext{ The simulated spectrum in Fig.~\ref{fig: ES}d) shows six dominant transition lines merging at $\nu\approx 0$ as $\epsilon \to 0$, which is a direct signature of the expected four-fold degeneracy of the ground state of $\tilde{H}_A$ in the thermodynamic limit (in the zero magnetization sector)}%~\footnote{Note that -- in contrast to the traditional spin-1 Haldane chain, which exhibits a two-fold degeneracy -- we expect a four-fold degeneracy due to the presence of an additional decoupled spin-1/2 at the bottom right corner.}
. Finally, we sweep $\lambda$ from \btext{$0$ to $1$}, as illustrated in Fig.~\ref{fig: ES}e), where the structure of resonant peaks shifts to $\nu=0$. This directly reflects the expected changes of the ES [cf. Fig.~\ref{fig: ES}b)], demonstrating that the on-device entanglement spectroscopy enables us to probe the transition from the trivial to the topological phase.

\emph{Outlook -- } Quantum variational learning provides a universal experimental toolset in the ongoing experimental effort to characterize novel equilibrium and non-equilibrium quantum phases \cite{semeghini2021probing} via their entanglement structure. %While entanglement data obtained in the present framework can be used to train machine learning algorithms to identify quantum phases \cite{van2017learning}, the protocol \textit{per se} has also immediate extensions e.g.~as \textit{on device} preparation and analysis of EH eigenstates \cite{larose2019variational}.  
Entanglement data obtained in the present framework can serve as input for further classical analysis, e.g., to train machine learning algorithms to identify quantum phases \cite{van2017learning}.
The fact that the optimization is performed \textit{on device} is a key feature of our protocol, which not only enables the subsequent spectroscopy, but also provides robustness against potential miscalibration of the experimental setup. \btext{Additionally, since our cost function is built from local observables, we expect the optimization to behave favorably under the barren plateau problem \cite{barren_plateau}, though further investigations are required.} 
%Moreover, the protocol \textit{per se} has also immediate extensions e.g.~as \textit{on device} preparation and analysis of EH eigenstates \cite{larose2019variational}.  

\textit{Acknowledgment --}
We thank L.\ K. Joshi, R.~Kaubr\"ugger, J.\ Carrasco, J.\ Yu, and B.\ Kraus for valuable discussions. We thank Ana Maria Rey and Murray Holland for a careful reading of the manuscript. We acknowledge funding from the European Union's Horizon 2020 research and innovation programme under Grant Agreement No.\ 817482 (Pasquans) and No.\ 731473 (QuantERA via QT-FLAG). Furthermore, this work was supported by the Simons Collaboration on Ultra-Quantum Matter, which is a grant from the Simons Foundation (651440, P.Z.), and LASCEM by AFOSR No. 64896-PH-QC. M.D.\ is partly supported by the ERC under grant number 758329 (AGEnTh). A.E.\ acknowledges funding by the German National Academy of Sciences Leopoldina under the grant number LPDS 2021-02. BV acknowledges funding from the Austrian Science Fundation (FWF, P 32597 N), and the French National Research Agency (ANR-20-CE47-0005, JCJC project QRand). The computational results presented here have been achieved (in part) using the LEO HPC infrastructure of the University of Innsbruck. 

\appendix
\counterwithin{figure}{section}

\section{Technical description of the quantum algorithm } \label{app:alg}
In the following, we first provide a technical description of our algorithm and then give a formal proof under which sufficient conditions the algorithm is guaranteed to succeed. We also discuss how symmetries of the reduced state affect the quantum variational learning of the EH. Further below, we apply the proof to the examples considered in the main text. 

\paragraph{Description of the protocol --}
In the first step of our protocol, we choose a variational class of Hamiltonians $\hat{H}^{\text{var}}_A(\boldsymbol{g})$ that can be realized on the quantum simulator. As discussed in the main text, the EH can often be well approximated by a set of variational parameters $\{g^\text{opt}_j\}$ that correspond to a deformation of the system Hamiltonian.

The main part of our algorithm finds such an optimal set of variational parameters in the following quantum-classical feedback loop:
\begin{enumerate}
\item[1.] Prepare an initial state of interest $\hat{\rho}$ at time $t=0$.
\item[2.] Evolve the reduced state $\hat{\rho}_A$ of the desired subsytem $A$ [see Eq.~\eqref{eq:reduced_density_operator}] with the ansatz $\hat{H}^{\text{var}}_A(\boldsymbol{g})$ up to some time $t>0$. Here we assume that the rest of the system does not interfere with the evolution in $A$. 
\item[3.] Measure the expectation value $\langle\hat{\mathcal{O}}_A\rangle_t$ [see Eq.~\eqref{eq:constraint}] of a set of observables $\{\mathcal{O}_A\}$ for several times $\{t\}$.
\item[4.] Calculate a cost function $\mathcal{C}(\boldsymbol{g})$ [see below and Eq.~\eqref{eq:cost_function}].
\item[5.] Repeat steps 2 to 5 for different variational parameters and minimize $\mathcal{C}(\boldsymbol{g})$.
\end{enumerate}

The cost function $\mathcal{C}$ is not unique, but we require the following property whose importance will become clear in the proof of the protocol: $(*)$ For sufficiently many observables $\{\mathcal{O}_A\}$ and observation times $\{t\}$, the minimization $\mathcal{C}(\boldsymbol{g}^\text{opt}) = \min_{\boldsymbol{g}} \left[ \mathcal{C}(\boldsymbol{g})\right]$ implies $\left[\hat{H}^{\text{var}}_A(\boldsymbol{g}^\text{opt}), \hat{\rho}_A \right] = 0$. A possible cost function, which we have used for the explicit calculations shown in the main text, is given by summing the squared deviations of the observables, at various times $t$, from their respective initial values,
\begin{align}\label{eq:cost_function}
    \mathcal{C}(\boldsymbol{g}) = \sum_{\mathcal{O}_A, t} \left[\langle \hat{\mathcal{O}}_A\rangle_t - \langle \hat{\mathcal{O}}_A\rangle_0 \right]^2 \;.
\end{align}
The desired property $(*)$ follows from the equations of motion for a set of observables that form a basis for all hermitian operators, monitored at all times $t>0$.

As illustrated in the main text, in practice few or even a single observable can be sufficient for a successful learning of the EH. We attribute this to the assumed local structure of the EH. Given that a local operator $\mathcal{O}_A$ will quickly scramble under the quantum dynamics generated by a generic many-body Hamiltonian $\hat{H}(\boldsymbol{g})$, one can expect that for the cost function given in Eq.~\eqref{eq:cost_function} the condition $\mathcal{C}(\boldsymbol{g}) = 0$ already provides sufficiently many constraints to conclude that $\left[\hat{H}^{\text{var}}_A(\boldsymbol{g}^\text{opt}), \hat{\rho}_A \right] = 0$.

\paragraph{Proof of the protocol --}
	Consider two commuting operators $\hat{X} = \sum_{j} x_j \hat{h}_j$ and $\hat{Y} = \sum_j y_j \hat{h}_j$. In particular, to prove our algorithm we have $\hat{X} = \hat{\tilde{H}}_A - c$ the exact EH (up to the normalization $c= \log \text{Tr}_A\left[\hat{X}\right]$) and $\hat{Y} = H^{\text{var}}_A (\boldsymbol{g}^\text{opt})$ the optimal variational ansatz. Then
	\begin{align}\label{eq:SM_commutator_zero}
		0 = \left[\hat{X},\hat{Y}\right] = \frac{1}{2} \sum_j\sum_{k\neq j} \left(x_j y_k - x_k y_j\right) \hat{C}_{jk} \;,
	\end{align}
	where we abbreviated $\hat{C}_{jk} = \left[\hat{h}_j, \hat{h}_k\right]$. $(i)$ If the collection of non-vanishing $\hat{C}_{jk}$ (excluding $\hat{C}_{kj} = -\hat{C}_{jk}$ to avoid double-counting) forms a set of linearly independent operators, then Eq.~\eqref{eq:SM_commutator_zero} implies the constraint $x_jy_k = x_k y_j$ for all pairs of coefficients corresponding to non-vanishing $\hat{C}_{jk}$. $(ii)$ If additionally the operators $\hat{h}_j$ are such that the non-vanishing $\hat{C}_{jk}$ provide a network of ``connected'' constraints, we can extend the equations $ x_j y_k = x_k y_j$ to \emph{all} coefficients. The solutions to this set of equations are given by $ x_j = \beta y_j$ with a single parameter $\beta$ for all $j$. In summary, given the two conditions $(i)$ linear independence of the $\hat{C}_{jk}$, and $(ii)$ ``connectedness'' of the $\hat{C}_{jk}\neq \hat{0}$, for the two operators $\hat{X}$ and $\hat{Y}$, we have shown that $\left[\hat{X},\hat{Y}\right] = 0$ already implies $\hat{X} = \beta \hat{Y}$.
	
	Given a tentative EH $\hat{\tilde{H}}_A = \sum_j g_j^{\text{opt}}\hat{h}_j + c$ and a corresponding ansatz $\hat{H}_A^\text{var}(\boldsymbol{g}) = \sum_j g_j\hat{h}_j$, we can thus check the applicability of our protocol, assuming a cost function that satisfies the property $(*)$, by calculating the commutator $\left[\hat{\tilde{H}}_A,\hat{H}^\text{var}_A(\boldsymbol{g})\right]$. Repeating the above considerations for a specific ansatz, it is straightforward to determine whether and to which extent a vanishing commutator relates the optimal ansatz with the exact EH. We demonstrate this procedure explicitly for the examples considered in this letter further below
	
	Following a similar  reasoning as put forward in \cite{qi2019determining}, we expect generic physical Hamiltonians of non-integrable models with quasi-local interactions and no further symmetries to fulfill the two necessary conditions. Intuitively, the quasi-locality directly provides the required ``connectedness'' together with a sufficiently strong restriction on allowed terms $\hat{h}_j$ such that the resulting commutators $\hat{C}_{jk}$ are generally independent.
	
	\paragraph{The role of symmetries --}
	Our protocol only determines relative strengths of coupling terms which are sensitive to the commutator with the EH. As a consequence all operators that commute with $\hat{\tilde{H}}_A$ lead to additional undetermined coefficients, which can be interpreted as generalized chemical potentials for the Gibbs state $\hat{\rho}_A = e^{-\hat{\tilde{H}}_A}$. There are at least two such coefficients, namely the prefactor $\beta$ corresponding to the conservation of ``energy'' of the EH, $\left[\hat{\tilde{H}}_A,\hat{\tilde{H}}_A\right] = 0$, and the constant $c$ fixed by the normalization $\text{Tr}_A \left[\hat{\rho}_A\right] = 1$. %These free constants can be seen as consequences of $\left[\hat{\tilde{H}}_A,\hat{\tilde{H}}_A\right] = 0$ and $\left[\hat{\tilde{H}}_A,\hat{\mathbf{1}}\right] = 0$, respectively. 
	Depending on the structure of the EH, there might be further symmetries corresponding to conserved charges $\hat{Q}$ with $\left[\hat{\tilde{H}}_A,\hat{Q}\right] = 0$, which result in further undetermined constants $\mu_Q$. Note that the assumption of a quasi-local EH implicitly excludes chemical potentials for symmetries which are not generated by an operator $\hat{Q} = \sum_{j_Q} \hat{q}_{j_Q}$ with quasi-local terms $\hat{q}_{j_Q}$. 
	
	For brevity, we next discuss the case of a single symmetry in more detail. Our considerations directly generalize to multiple (not necessarily global) symmetries if the corresponding conserved charges commute among each other, including the case of abelian gauge theories. Without loss of generality, we assume that the $\hat{q}_{j_Q}$ are included in the EH $\hat{\tilde{H}}_A$ and in the ansatz $\hat{H}^\text{var}_A(\boldsymbol{g})$, i.e. there exist $j=j_Q$ such that $\hat{h}_{j_Q} = \hat{q}_{j_Q} $. We can then uniquely rewrite 
	\begin{align}
		\hat{\tilde{H}}_A= \dots +  \sum_{j_Q} \delta 
		g^{\text{opt}}_{j_Q} \hat{h}_{j_Q} + \mu_Q \hat{Q} \:,
	\end{align}
	where we omitted terms that do not appear in $\hat{Q}$ and we introduced $\mu_Q = \sum_{j_Q} g^{\text{opt}}_{j_Q}$ and $\sum_{j_Q} \delta g^{\text{opt}}_{j_Q} =0$. With the additional constraint $\sum_{j_Q} g_{j_Q} =0$ imposed for $\hat{H}^\text{var}_A(\boldsymbol{g})$, the optimal solution can be written as
	\begin{align}
	    \hat{H}^\text{var}_A(\boldsymbol{g}^{\text{opt}}) = \frac{1}{\beta}(\hat{\tilde{H}}_A - c - \mu_Q \hat{Q})  \;,
	\end{align}
	i.e. our protocol uniquely determines the EH up to the ``global'' constants  $\beta, c$ and $\mu_Q$.
	
	Now, since $\left[\hat{\rho}_A, \hat{Q} \right] = 0$, there exists a basis where $\hat{\rho}_A = \sum_{\alpha, Q} e^{-\xi_{\alpha,Q}} | \Phi^A_{\alpha,Q} \rangle \langle \Phi^A_{\alpha,Q}|$ and we can separate $\xi_{\alpha,Q} = \xi'_{\alpha,Q} + \mu_Q Q$ with $\xi'_{\alpha,Q}$ the eigenvalues of $\hat{\tilde{H}}_A$ for $\mu_Q=0$ in a sector with fixed $Q$. Finally, we obtain that the probability $p_Q = \text{Tr} \left[\hat{\rho}_A \delta_{\hat{Q},Q} \right]$ to measure a particular value $Q$ is given by $p_Q = e^{-\mu_Q Q} p'_Q$ with $p'_Q = \sum_\alpha e^{-\xi'_{\alpha, Q}}$.
	
	These facts suggest to carry out our protocol in each $Q$-sector separately and first obtain the individual spectra $\left\lbrace \xi'_{\alpha,Q} \right\rbrace$. We emphasize that at this point we require knowledge of the correct values of $\beta$ and $c$ to fix the total scales. If one is interested in comparing different sectors, one can then calculate $p'_Q$ and measure $p_Q$ in order to deduce $\mu_Q$ from $e^{-\mu_Q Q} = p_Q / p'_Q$. This approach is feasible in cases where the operator $\hat{Q}$ can be measured in single shots of an experiment, such that the data can be sorted according to the observed values of $Q$. This is the case for the examples considered in this letter where $Q$ corresponds to magnetization or particle number. 
	Resolving different symmetry sectors becomes particularly important in the context of quantum Hall states, where Li and Haldane conjectured~[] that the momentum dependence of the ES can reveal the structure of the relevant CFT governing gapless edge excitations.

	\paragraph{Hubbard model --} For the Hubbard model, we assume that the exact EH has the form
	\begin{align}
		\hat{\tilde{H}}_A = &-\sum_{\langle nm \rangle, \sigma } t_{nm}\hat{T}^\sigma_{nm} + \sum_n U_n \hat{n}^\uparrow_n \hat{n}^\downarrow_m \nonumber \\&+ \sum_{n,\sigma} \delta \mu_n^\sigma \hat{n}^\sigma_n + \sum_\sigma \mu^\sigma \hat{N}^\sigma + c \;,
	\end{align}
	with $t_{nm} = t_{mn}$ and $\sum_n \delta \mu^\sigma_n=0$. Here, we abbreviated the hopping terms $\hat{T}^\sigma_{nm} = \hat{c}^\dagger_{n,\sigma}\hat{c}_{m,\sigma} + \text{h.c.}$, the densities $\hat{n}^\sigma_n = \hat{c}^\dagger_{n,\sigma}\hat{c}_{n,\sigma}$ and the total particle numbers $\hat{N}^\sigma = \sum_n\hat{n}^\sigma_n$. This EH conserves total particle number $\hat{N} = \hat{N}^\uparrow + \hat{N}^\downarrow$ and the total magnetization  $\hat{M} =  \hat{N}^\uparrow - \hat{N}^\downarrow $, i.e. $\left[	\hat{\tilde{H}}_A, \hat{N}  \right] = \left[	\hat{\tilde{H}}_A, \hat{M}  \right] = 0$ and therefore $\hat{\tilde{H}}_A$ parametrizes a reduced density matrix that is block diagonal w.r.t. $\hat{N}$ and $\hat{M}$.
	
	To check whether $\hat{\tilde{H}}_A$ can be learned by our protocol, we make an ansatz $\hat{H}_A^\text{var}$ of the same form with free parameters $\tilde{t}_{nm}$, $\tilde{U}_n$, $\delta\tilde{\mu}^\sigma_n$, but $\sum_n\delta\tilde{\mu}^\sigma_n = 0$, and calculate the commutator $\left[\hat{\tilde{H}}_A,\hat{H}_A^\text{var}\right]$, which involves the following three types of non-vanishing commutators,
	\begin{widetext}
	\begin{subequations}
	\begin{align}
				\left[\hat{T}^\sigma_{nm},\hat{T}^\rho_{n'm'}\right] &= \delta^{\sigma \rho} \delta_{mn'}\left[ \left(\hat{c}^\dagger_{n,\sigma} \hat{c}_{m',\sigma} - \text{h.c.}\right) + \delta_{nm'}\right] + \delta^{\sigma \rho} \delta_{mm'}\left[ \left(\hat{c}^\dagger_{n,\sigma} \hat{c}_{n',\sigma} - \text{h.c.}\right) + \delta_{nn'}\right] \nonumber \\ &\quad+
				 \delta^{\sigma \rho} \delta_{nn'}\left[ \left(\hat{c}^\dagger_{m,\sigma} \hat{c}_{m',\sigma} - \text{h.c.}\right) + \delta_{mm'}\right] + \delta^{\sigma \rho} \delta_{nm'}\left[ \left(\hat{c}^\dagger_{m,\sigma} \hat{c}_{n',\sigma} - \text{h.c.}\right) + \delta_{mn'}\right] \;,\\
		\left[\hat{T}^\sigma_{nm}, \hat{n}^\rho_{n'}\right] &= \delta^{\rho \sigma} \left(-\delta_{nn'} + \delta_{mn'}\right) \left(\hat{c}^\dagger_{n,\sigma} \hat{c}_{m,\sigma} - \text{h.c.}\right) \;, \\
			\left[\hat{T}^\sigma_{nm}, \hat{n}^\uparrow_{n'} \hat{n}^\downarrow_{n'}\right] &= \delta^{\sigma \uparrow} \left(\hat{c}^\dagger_{n,\uparrow} \hat{c}_{m,\uparrow} - \text{h.c.} \right)\left(-\delta_{nn'} \hat{n}^\downarrow_n + \delta_{mn'} \hat{n}^\downarrow_m\right) + \delta^{\sigma \downarrow} \left(\hat{c}^\dagger_{n,\downarrow} \hat{c}_{m,\downarrow} - \text{h.c.} \right)\left(-\delta_{nn'} \hat{n}^\uparrow_n + \delta_{mn'} \hat{n}^\uparrow_m\right) \;.
	\end{align}
\end{subequations}
\end{widetext}

	Collecting the coefficients of linearly independent operators in the equation $\left[\hat{\tilde{H}}_A,\hat{H}_A^\text{var}\right] = 0$, we obtain three types of constraints. First, there are terms proportional to fermion bilinears of the form $\left(\hat{c}^\dagger_{n,\sigma} \hat{c}_{\ell,\sigma} - \text{h.c.}\right)$ with support on two different sites $n$ and $\ell$ which share a common neighbour $m$. For such indices, we obtain the constraints
	\begin{align}
		t_{nm}\tilde{t}_{m\ell}  = 	\tilde{t}_{nm} t_{m\ell} \;.
	\end{align}
	Similarly, terms involving four fermion operators, such as $ \left(\hat{c}^\dagger_{n,\uparrow} \hat{c}_{m,\uparrow} - \text{h.c.} \right) \hat{n}^\downarrow_n $ for nearest neighbours $n$ and $m$ lead to the constraints
	\begin{align}
		\tilde{t}_{nm} U_{n} = \tilde{U}_n t_{nm} \;.
	\end{align}
	Solving these sets of constraints, we obtain $\beta \tilde{t}_{nm} =  t_{nm}$ and $\beta\tilde{U}_{nm} =  U_{nm}$ with arbitrary $\beta$. Third, the remaining linearly independent operators are of the form $\left(\hat{c}^\dagger_{n,\sigma} \hat{c}_{m,\sigma} - \text{h.c.}\right)$ supported on nearest neighbours $n$ and $m$. Using the relation between $t$ and $\tilde{t}$, the remaining coefficients yield the constraints
	\begin{align}
		\beta\left(\delta \mu^\sigma_m - \delta \mu^\sigma_n \right) = \delta \tilde{\mu}^\sigma_m - \delta \tilde{\mu}^\sigma_n \;.
	\end{align}
	Under the condition $\sum_n \delta \mu^\sigma_n= \sum_n \delta \tilde{\mu}^\sigma_n = 0$, the solution is given by $\beta \tilde{\mu}^\sigma_{n} =  \mu^\sigma_{n}$. In summary, we have $\hat{H}_A^\text{var} = \frac{1}{\beta} \left( \hat{\tilde{H}}_A - c - \sum_\sigma \mu^\sigma \hat{N}^\sigma\right)$, i.e. the EH is determined up to the four constants $\beta$, $c$, $\mu^\uparrow$ and $\mu^\downarrow$.

\section{Realizing time evolution with the deformed Fermi Hubbard Hamiltonian as digital quantum simulation} \label{app:trotter}
% \subsection{Realizing Deformations of the Fermi-Hubbard model using Trotter sequenzes}
	
Quantum variational learning of the EH relies on the ability to time evolve a subsystem $A$ with a spatially deformed Hamiltonian [see Eq.~(\ref{eq:constraint}) in main text]. In the case of the FHM [see Eq.~(\ref{eq:FHM}) in main text] physically realizing these deformations as independent local control of $J_{jk},U,\mu_\sigma$ may be challenging.
 Here we discuss strategies how to realize time evolution with the deformed Fermi Hubbard Hamiltonian as digital quantum simulation, i.e.~as a series of Trotter steps representing  consecutive stroboscopic quantum quenches. As we discuss below, this approach minimizes the experimental requirements for the local controllability of the onsite interaction energies. Simultaneously, the scheme provides a natural realization of the required deformations %required for our Hamiltonian learning protocol
$\hat{H}_A^\text{var}(\boldsymbol{g}) = \sum_i g_i \hat{h}_i$, with $\hat{h}_i$ the fundamental Hamiltonian building blocks as discussed in the main text.
	
\begin{figure*}[t]
\includegraphics[width=0.95\textwidth] {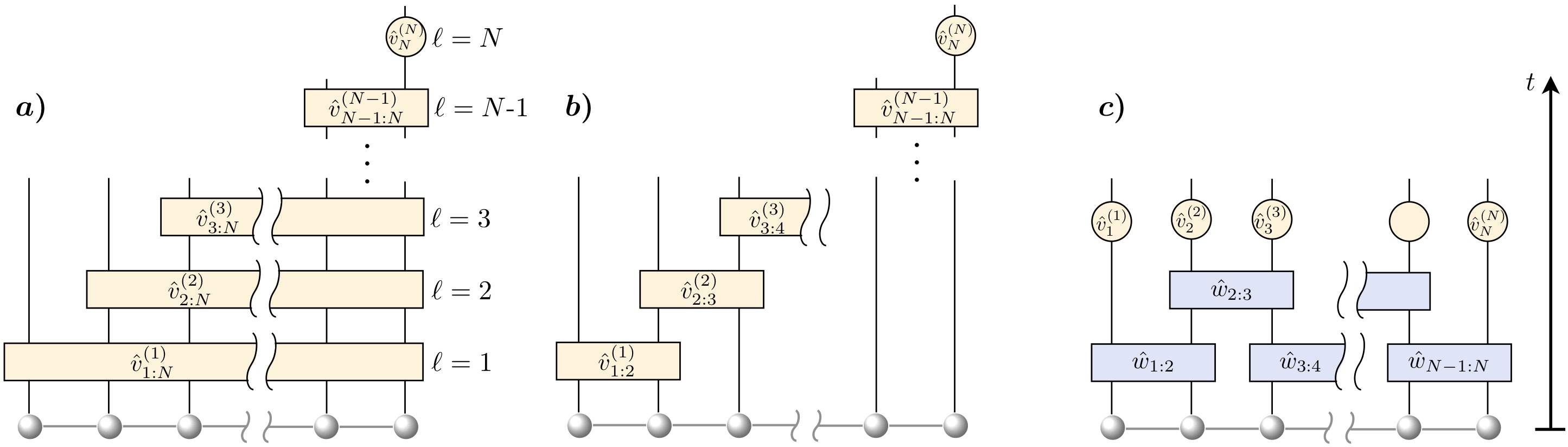}
\caption{\textit{Different schemes for implementing a single Trotter time step with a spatially deformed Hubbard Hamiltonian} a) A single Trotter time step given in Eq.~(\ref{eq:def}) assembled from building blocks acting on gradually shrinking subsystems of a 1D lattice. The spatial dependence of the interaction energies $U_i$ and the chemical potentials $\mu_i$ is achieved by accumulating phases during time evolution with the unitary building blocks  $\hat{v}_{A'}^{\ell}$. b) Like a) but with unitary building blocks acting on two neighbouring lattice sites. c) Implementation motivated by TEBD with local beam-splitting operations defined in Eq.~(\ref{eq:beamsplitter}). }
\label{fig:trotter}
\end{figure*}

In the remainder of this section we discuss the realization of a single Trotter time step with a deformed Hamiltonian described by spatially varying parameters $J_{i, i+1}$, $U_i$ and $\mu_i$. %In order to not overburden the notation, 
For notational clarity, we discuss the implementation for 1D and comment on extensions to higher dimensional systems at the end of this section. In the 1D case, the unitary operation implementing a single Trotter time step with a deformed Hubbard Hamiltonian reads
\begin{widetext}
\begin{align} \label{eq:def}
e^{-i \hat{H}^\text{var} \tau}  = \exp \left\{ -i \tau \left[  -\sum_{i\sigma} J_{i, i+1} \left( \hat{c}_{i \sigma}^{\dagger} \hat{c}_{i+1\, \sigma} + \text{H.c.} \right) + \sum_i U_i \hat{n}_{i \uparrow} \hat{n}_{i \downarrow}  - \sum_{i \sigma} \mu_i \hat{n}_{i \sigma}  \right]\right\} = \hat{V}(\tau,  \{J_{ij}, U_i, \mu_i\})  + \mathcal{O}(\tau^2) .
\end{align}
\end{widetext}
%With the operator specified in Eq.~(\ref{eq:def}), 
The time evolution up to time $T = n \tau$ can be approximated as $e^{-i \hat{H}^\text{var} T} \simeq \hat{V}^n(\tau,  \{J_{ij}, U_i, \mu_i\})$ up to Trotter errors depending on the size of the time step $\tau$.

As sketched in Fig.~\ref{fig:trotter}, the unitary given in Eq.~(\ref{eq:def}), can be assembled from smaller building blocks, consecutively acting on different regions $A'$ of the sublattice. %As can be seen, 
The circuits in panel a) and b) exhibit a layered structure, $V = \prod_\ell v^{(\ell)}_{A'}$, where the unitary operations %$\hat{v}_{A'}^{(\ell)}$ 
\begin{widetext}
\begin{align} \label{eq:vA}
\hat{v}_{A'}^{(\ell)}(J^{(\ell)}, T^{(\ell)}) = \exp \left \{ -i T^{(\ell)} \left[ -J^{(\ell)} \sum_{i \in A', \sigma}  \left( \hat{c}_{i \sigma}^{\dagger} \hat{c}_{i+1\, \sigma} + \text{H.c.}  \right) + U \sum_{i \in A'} \hat{n}_{i \uparrow} \hat{n}_{i \downarrow}  - \mu \sum_{i \in A', \sigma} \hat{n}_{i \sigma} \right] \right\}
\end{align}
\end{widetext}
at a given layer $\ell$
depend on two parameters: %given by 
a uniform tunnelling matrix element $J^{(\ell)}$ and the quench time $T^{(\ell)}$ measured in units of the Trotter time step $\tau$.

Assembling the Trotter time step Eq.~(\ref{eq:def}) from building blocks given in Eq.~(\ref{eq:vA}) comes with several advantages. First, %note that 
the onsite interaction $U$ and the chemical potential $\mu$ are independent of the current layer $\ell$, thus exhibiting a constant value throughout all building blocks $\hat{v}_{A'}^{(\ell)}$.  Second, the only Hamiltonian parameter to be controlled in $\hat{v}_{A'}^{(\ell)}$ is a global tunneling amplitude $J^{(\ell)}$, while the spatial dependence of the parameters $U_i$ and $\mu_i$ [Eq.~(\ref{eq:def})] is eventually achieved by accumulating phases from varying the quench times for the current layer $\ell$. This is particularly useful for deformations of the BW-type, where %for a subsystem in a bipartite lattice
the EH parameters follow a ramp, reaching their maximum value at the boundary. 
	
For a sufficiently small Trotter time step $\tau$, it is straightforward to relate the building block parameters $J^{(\ell)}$ and $T^{(\ell)}$ to the final deformation parameters $J_{i, i+1}$, $U_i$ and $\mu_i$ is p. For the circuit displayed in Fig.~\ref{fig:trotter} a), summing up all contributions from each layer together with $J^{(N)} = 0$ we obtain %the relation is approximately given by
\begin{align} \label{eq:Jii}
J_{i, i+1} &\simeq \sum_{\ell = 1}^{i}  J^{(\ell)} T^{(\ell)} \;, \\ \label{eq:Ui}
U_i &\simeq U \sum_{\ell = 1}^{i}  T^{(\ell)} \;,\\ \label{eq:mui}
\mu_i &\simeq \mu \sum_{\ell = 1}^{i}  T^{(\ell)} \;,
\end{align}
%which is obtained by summing up all contributions from each layer with $J^{(N)} = 0$.
which becomes exact in the limit $\tau \rightarrow 0$.
Eq.~(\ref{eq:Jii})-(\ref{eq:mui}) provide a system of equations with a unique solution for the unknown control parameters $J^{(\ell)}$ and $T^{(\ell)}$. However, from Eq.~(\ref{eq:Jii})-(\ref{eq:mui}) it is evident that the parameters $U_i$ and $\mu_i$ cannot be controlled independently since they obey a fixed ratio $U_i/\mu_i = U/\mu$. Crucially, this restriction exactly matches the type of deformations required for the ansatz $\hat{H}_A^\text{var}(\boldsymbol{g}) = \sum_{i} g_i \hat{h}_i$ since the ratio $U_i/\mu_i$ is constant in $\hat{h}_i$. 

The structure of the circuits depicted in Fig.~\ref{fig:trotter} a) and b) facilitates an iterative solution of the system of equations relating the parameters $J^{(\ell)}$ and $T^{(\ell)}$ to the final deformation parameters. Given a deformation pattern $\{ J_{i, i+1}, U_i\}$, the solution for the case of Fig.~\ref{fig:trotter} a) reads 
\begin{align} \label{eq:Jell}
J^{(\ell)} &= U \frac{J_{\ell, \ell+1} - \sum_{i = 1}^{\ell - 1} (-1)^{i-1} U_{\ell - i}}{U_{\ell} - \sum_{i = 1}^{\ell - 1} (-1)^{i - 1} U_{\ell - 1}} \\ \label{eq:Uell} \nonumber \\[-2ex] 
T^{(\ell)} &= \frac{1}{U} \left[ U_{\ell} - \sum_{i = 1}^{\ell - 1} (-1)^{i - 1} U_{\ell - 1} \right]. 
\end{align}
The solution for the circuit in panel b) exhibits an analogous structure, with the only difference that the numerator in Eq.~(\ref{eq:Jell}) is  given by $J_{\ell, \ell+1}$. Note that in these two examples, the number of layers coincides with the number of lattice sites in the subsystem.

Finally, we briefly discuss a third version for realizing a single Trotter time step [see Fig.~\ref{fig:trotter} c)] which is reminiscent of the time-evolving block decimation (TEBD) with matrix product states \cite{PhysRevLett.91.147902}. Here, each building block depends on a single parameter given by the time window of the local quenches. The blocks labeled by $\hat{w}_{i, i+1}$ act as local beam-splitter operations on neighbouring lattice sites given by
\begin{align} \label{eq:beamsplitter}
\hat{w}_{i, i+1} = \exp \left[i T^{(i, i+1)} J \sum_{i \sigma}  \left( \hat{c}_{i \sigma}^{\dagger} \hat{c}_{i+1\, \sigma} + \text{H.c.}  \right) \right].
\end{align}
In this case the deformation parameters are directly related to the quench times $J_{i, i+1} = T^{(i, i+1)} J$, $U_i = U T^{(i)}$ and $\mu_i = \mu T^{(i)}$. 

The schemes described above can be extended to 2D and higher-dimensional lattices in a straight-forward way. Like in the 1D case, the general idea is to decompose the Trotter time step into smaller building blocks which act on different subregions of the lattice, each depending on a certain number of control parameters. The total number of control parameters determines the flexibility of the variational ansatz and the parameters of the building blocks can be related to the final deformation parameters via a system of linear equations.

\section{ Learning the EH by minimizing the relative entropy} \label{app:relent}
In Fig.~\ref{fig:FermiHubbard}, we showed that the $g_j$ parameters obtained by our variational optimization agree well with the optimal EH that is obtained by minimizing the relative entropy~\cite{PhysRevLett.122.150606}. Here we describe the latter method. We note that minimizing the relative entropy gives the optimal EH including the scale factor $\beta$ and the overall shift $c$.

The reduced density matrix for the EH ansatz $\tilde{H}_A$ is $\tilde{\rho}_A = \exp(-\tilde{H}_A)/{\rm Tr}( \exp(-\tilde{H}_A))$. The statistical distinguishability of $\tilde{\rho}_A$ and the true density matrix $\rho_A$ can be quantified by the quantum relative entropy $S(\rho_A \vert\vert \tilde{\rho}_A) = {\rm Tr} \left(\rho_A (\log\rho_A - \log\tilde{\rho}_A)\right)$. In terms of $\tilde{H}_A$, we have
\begin{equation} \label{eqn: rel entropy}
S(\rho_A \vert\vert \tilde{\rho}_A) = -S_0 + {\rm Tr}(\rho_A \tilde{H}_A) + \log {\rm Tr}( \exp(-\tilde{H}_A)),
\end{equation}
where $S_0 = -{\rm Tr}\rho_A \log\rho_A$ is the von Neumann entropy of $\rho_A$.

It can be shown that $S(\rho_A \vert\vert \tilde{\rho}_A) \geq 0$, with equality iff $\rho_A = \tilde{\rho}_A$. Thus, one can learn an optimal approximation of the EH by minimizing $S(\rho_A \vert\vert \tilde{\rho}_A)$. For the minimization, the first term in Eq.~\eqref{eqn: rel entropy} is an irrelevant constant. The second term is the expectation value of $\tilde{H}_A$, which requires to compute numerically or measure in an experiment the components of the given ansatz. The third term plays the role of a free energy, ensuring the correct normalization, and effectively corresponds to the constant $c$.

In Fig.~\ref{fig:FermiHubbard} of the main text, the dashed lines labelled by ``optimal EH'' correspond to the EH parameters found by minimizing the relative entropy as described above. In order to compare the optimal parameters of our QVL protocol to the optimal parameters of the relative entropy minimization, we have rescaled the former by a single constant $\beta$. For this comparison, we have used the same constant as for the analysis presented in Fig.~\ref{fig:Fid}. Namely, $\beta$ is obtained independently by maximizing the fidelity of a thermal state corresponding to EH (with the deformation fixed by QVL) at inverse temperature $\beta$  w.r.t. the exact reduced density operator.

\section{Determining the constants $\beta$ and $c$ from a scaling-hypothesis} \label{app:scale}

\begin{figure}[t]
\includegraphics[width=1.0\columnwidth] {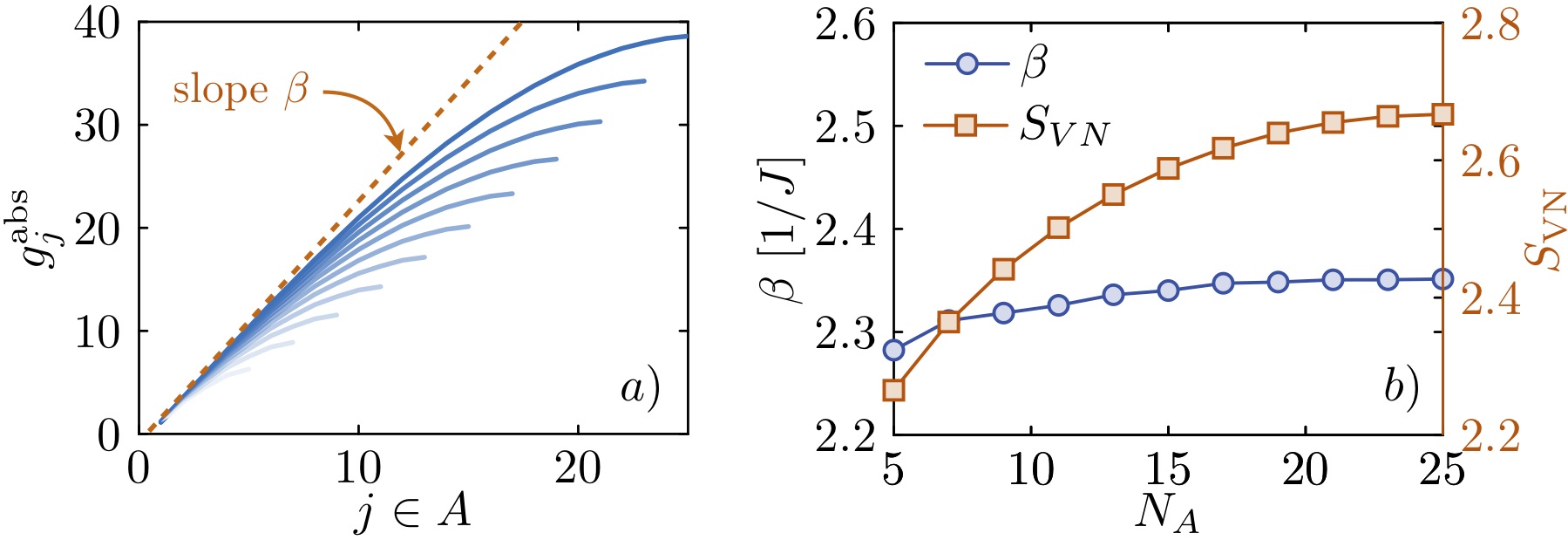}
\caption{\textit{BW ramps and $\beta$-prefactor as a function of the subsystem size in a 1D Hubbard chain.} a) Rescaled coefficients of the EH for different bipartitions of a 50-site Hubbard chain with $U = 1$ and $J = 1$ determined from entanglement eigenstates (see main text). b) Initial slope of the BW gradient $\beta$ determined from panel a) and von Neumann entropy as a function of the subsystem size. 
}
\label{fig:beta}
\end{figure}
Quantum variational learning of the EH, as described in the main text, determines the EH in terms of the coefficients $g_j$ up to a global prefactor $\beta$ and the normalization constant $c$. Below we discuss a possible strategy to infer this missing information by combining the results of our protocol with additional properties, extracted for small subsystems. In particular, the universality of the BW deformation suggests to determine the missing prefactor $\beta$ by matching a common slope to the optimal parameters $g_j^\text{opt}(N_A)$ for different sizes $N_A$ of the subsystem $A$. This would enable to determine $\beta$ for large subsystem sizes by extrapolating the prefactor from smaller subsystems where it can be determined by other means, e.g. by tomographic methods \cite{kokail2020entanglement, choo2018measurement}. Alternatively, one can learn $\beta$ more efficiently by minimizing the relative entropy of a normalized state $\propto \exp (-\beta \hat{H}^\text{var}_A)$ w.r.t. the exact density operator $\hat{\rho}_A$, which requires a classical diagonalization of $\hat{H}^\text{var}_A$, but only a measurement of $\text{Tr} \left[\hat{\rho}_A \, \hat{H}^\text{var}_A\right]$.  %determined by tomographic methods \alert{cite: NatPhys, CompSens...}.

In the following we verify that such an extrapolation is possible by analysing the EH for subsystems on the right boundary of a 50-site Hubbard chain with $U/J = 1$. To this end, we first perform DMRG calculations to obtain an %highly 
accurate MPS representation of the Hubbard ground state. Second, we use a method %for determining 
to infer the structure of the EH from entanglement eigenstates $\ket{\Phi_A^{\alpha}}$ \cite{zhu2019reconstructing} for different subsystem sizes $N_A$ on the right boundary of the chain which results in a ``normalized'' EH $\hat{H}_A(\boldsymbol{g}) = \sum_{j \in A} g_j \hat{h}_j$ with coefficients $g_j$ normalized to one ($\sum_j g_j^2 = 1$). The correctly scaled EH $\hat{H}_A^{\text{abs}}$ that determines the reduced density matrix via $\hat{\rho}_A = \exp( - \hat{H}_A^{\text{abs}} )$ is approximately related to the normalized EH by a shift and a multiplicative factor $\hat{H}_A^\text{abs} \approx a + b \hat{H}_A$. We estimate the parameters $a$ and $b$  through a linear fit of the eigenvalues $\xi_{\alpha} = \bra{\Phi_A^{\alpha}} \hat{H}_A \ket{\Phi_A^{\alpha}} $ versus the Schmidt coefficients $\lambda_{\alpha}^2$ which are a natural byproduct of DMRG calculations: $-\log(\lambda_{\alpha}^2) \simeq a + b \xi_{\alpha}$. 

Having determined the fit parameters $a$ and $b$, we %allows us to
define the absolute EH parameters as $g_j^\text{abs} = b g_j$ which are plotted in Fig.~\ref{fig:beta} a). The factor $\beta$ is now defined as the derivative (finite difference) of the coefficients $g_j^\text{abs}$ at the left boundary of the subsystem where the curves in Fig.~\ref{fig:beta} a) intersect the $x$-axis. We determine the finite difference by fitting the BW ramps with a sin-function: $f(j, \boldsymbol{q}) = q_1 + q_2 \text{sin}[q_3 (x - q_4)]$ as suggested by CFT \cite{Cardy_2016}. As shown in Fig.~\ref{fig:beta} b), $\beta$ is largely independent of the subsystem size, which supports our proposed extrapolation. Given $\beta$, the constant $c$ follows from the normalization of the ES $\left(\sum_\alpha e^{-\xi_\alpha} = 1\right)$.

\section{Comparison between quantum and classical post-processing for learning the EH} \label{app:comp}

In this section we further elaborate on the relation of our quantum variational learning protocol to classical protocols in which the Hamiltonian is inferred from local measurements on the input state $\ket{\Psi_G}$. Additionally, we provide a detailed description of the classical post-processing step used to obtain the data presented in Fig.~\ref{fig:Fid} in the main text, including a discussion of the required observables and constraints. Finally, we comment on possible modifications of our protocol in order to extend its applicability to a wider class of Hamiltonians.

\subsection{Relation to classical Hamiltonian learning protocols}
As discussed in the main text, the core element of our protocol is the minimization of time variations of observables $\hat{\mathcal{O}}_A$, with the aim of enforcing $\braket{\hat{\mathcal{O}}_A}_t = \text{const}$. By evolving the quantum state for a finite time, the quantum device effectively includes and evaluates all commutators of the deformed Hamiltonian $\hat{H}_A^\text{var}(\boldsymbol{g})$ with the chosen observables $\hat{\mathcal{O}}_A$, which follows from expanding the operators within the trace in Eq.~(\ref{eq:constraint}) of the main text as
\begin{widetext}
\begin{align}
\begin{split}
    e^{i \hat{H}_A^\text{var}(\boldsymbol{g})t} \hat{\mathcal{O}}_A e^{-i \hat{H}_A^\text{var}(\boldsymbol{g})t} &= \hat{\mathcal{O}}_A + it [\hat{H}_A^\text{var}(\boldsymbol{g}), \hat{\mathcal{O}}_A] \\
    &- \frac{t^2}{2} [\hat{H}_A^\text{var}(\boldsymbol{g}),[\hat{H}_A^\text{var}(\boldsymbol{g}),\hat{\mathcal{O}}_A]] -i \frac{t^3}{6} [\hat{H}_A^\text{var}(\boldsymbol{g}),[\hat{H}_A^\text{var}(\boldsymbol{g}),[\hat{H}_A^\text{var}(\boldsymbol{g}),\hat{\mathcal{O}}_A]]] + \cdots \ .
\end{split} \label{eq:commquant}
\end{align}
\end{widetext}
Our algorithm aims to enforce $\braket{\hat{\mathcal{O}}_A}_t = \text{const}$, which effectively implies that almost all commutators in Eq.~(\ref{eq:commquant}) vanish, and thereby provides a sufficient amount of (nonlinear) constraints to uniquely determine $\hat{H}_A^\text{var}(\boldsymbol{g})$ from a small number of simple observables $\hat{\mathcal{O}}_A$. In particular, in the examples presented in the main text we select local mutually commuting observables accessible in state-of-the-art cold atom experiments.

In contrast, classical Hamiltonian learning protocols as described in \cite{bairey2019learning, qi2019determining} are based on the first order in time expansion of Eq.~(\ref{eq:commquant}). The idea is to construct a variational Hamiltonian as an expansion in a local operator basis $\{ \hat{h}_j \}$: $\hat{H}_A^\text{var}(\boldsymbol{g}) = \sum_j g_j \hat{h}_j$ and to determine the coefficients $g_j$ in such a way that the first order commutator in Eq.~(\ref{eq:commquant}) vanishes for suitably chosen constraints $\hat{\mathcal{O}}_A^{(n)}$. Plugging the expansion of $\hat{H}_A^\text{var}(\boldsymbol{g})$ into the commutator results in a system of linear equations $M \boldsymbol{g} = \boldsymbol{0}$ with matrix elements given by $M_{jn} = i \braket{[\hat{h}_j, \hat{\mathcal{O}}_A^{(n)}]}$ evaluated in the initial state. Here the number of constraints $\hat{\mathcal{O}}_A$ needs to be larger or equal than the number of elements in the local operator basis $\{\hat{h}_j \}$ in order to have at least as many equations as unknowns. The optimal Hamiltonian coefficients $\boldsymbol{g}^\text{opt}$ are given by the singular vector of $M$ corresponding to the smallest singular value, thus minimizing the norm $\parallel \! M \boldsymbol{g} \! \parallel$.

In Fig.~\ref{fig:Fid} in the main text we compare classical learning of the EH as described in the previous paragraph to our variational protocol in which we use different parametrizations for $\hat{H}_A^\text{var}(\boldsymbol{g})$. The operators in the local basis $\{\hat{h}_j \}$ are chosen to be the same as the ones used in our deformed variational Hamiltonian [Eq.~(\ref{eq:hj}) main text]. For the constraints we select operators $\hat{\mathcal{O}}^{(n\sigma \sigma')} = i(\hat{c}^{\dagger}_{n \sigma} \hat{c}_{n+1 \, \sigma} - \text{H.c.}) \hat{n}_{j \pm 2 \,  \sigma'}$, which provide a sufficient number of linear independent rows in $M$ [see also Ref.~\cite{carrasco2021theoretical}]. In order to analyse the effect of a finite number of samples $N_M$, we impose Gaussian noise on the matrix elements $M_{jn}$ with a variance $\epsilon^2$ given by $\epsilon^2 = \text{Var} ([\hat{h}_j, \hat{\mathcal{O}}_A^{(n)}])/N_M$. As discussed in the main text, while the classical protocol is more flexible because it does not require a physical realization of a deformed Hamiltonian, it requires (for the example studied here) a higher number of experimental runs as well as measurements of observables which are more difficult to access experimentally.

\subsection{Extensions to a wider class of variational Hamiltonians}
In the main text, we have focused on cases where the EH is well approximated by a deformation of the system Hamiltonian. Since the exact form of the EH is not known for general quantum many-body states, it might be necessary to include more, possibly non-local, terms in the ansatz for our variational protocol. While this is straightforward for the classical learning protocol, our quantum variational learning protocol appears to be restricted to directly experimentally realizable Hamiltonians. However, as a natural extension consider time evolution generated by a more general Hamiltonian $\hat{K}_A^\text{var}(\boldsymbol{g}, \boldsymbol{\theta}) = U(\boldsymbol{\theta}) \hat{H}_A^\text{var}(\boldsymbol{g}) U^{\dagger}(\boldsymbol{\theta})$, unitarily related to the original deformation $\hat{H}_A^\text{var}(\boldsymbol{g})$. Since
\begin{align}
    U(\boldsymbol{\theta})e^{-i \hat{H}_A^\text{var}(\boldsymbol{g})t} U^{\dagger}(\boldsymbol{\theta}) = e^{-i \hat{K}_A^\text{var}(\boldsymbol{g}, \boldsymbol{\theta})t} \;,
\end{align}
we can effectively enlarge the variational class by including an additional ``encoding'' and ``decoding'' step determined by the unitary operator $U(\boldsymbol{\theta})$, parametrized by additional variational parameters $\boldsymbol{\theta}$. This extension in principle allows to include arbitrary additional terms in the variational ansatz, depending on the experimentally available unitary operations $U(\boldsymbol{\theta})$.

\section{Spectroscopy of the EH}
		
The inspiration for our protocol to measure the ES is off-resonant two-level Rabi oscillations. The wave function of a two-level system, with detuning $\Delta$ and Rabi coupling $\Omega$, undergoes oscillations at frequency $\sqrt{ \Omega^2 + \Delta^2/4}$. Observables oscillate at twice this frequency, where the factor of two speedup arises from squaring the amplitudes of the wavefunction. Thus, the detuning $\Delta$ can be obtained from the Fourier peaks of the measurements of any typical observable $\braket{ \hat{O}(t) }$, in the limit $\Omega \rightarrow 0$.

Our protocol to obtain universal ratios of the ES measures observables when the system is driven by $\hat{H}(\boldsymbol{g}^{\rm opt}) + \epsilon \hat{H}'$. Our system is multi-level, with the different levels being the Schmidt vectors. The first term, $\hat{H}(\boldsymbol{g}^{\rm opt})$, is diagonal in the Schmidt vector basis, and thus plays the role of detunings in the two-level example. The perturbation $\epsilon\hat{H}'$ has nonzero matrix elements between the Schmidt vectors, and thus provides the Rabi couplings $\Omega$. We chose the perturbation as an unequal magnetic field on two lattice sites, which breaks SU(2) and U(1) symmetries. Therefore, all Rabi couplings between every pair of Schmidt vectors is nonzero. Our system can then be essentially viewed as undergoing several off-resonant two-level Rabi oscillations \footnote{For degenerate levels, these will be resonant Rabi oscillations}. Note that the dynamics do not decompose into that of independent two-level systems, therefore the above two-level picture is not strictly valid, but the two-level picture still suffices to understand the essential physics.

\begin{figure}[t]
\includegraphics[width=1.0\columnwidth]{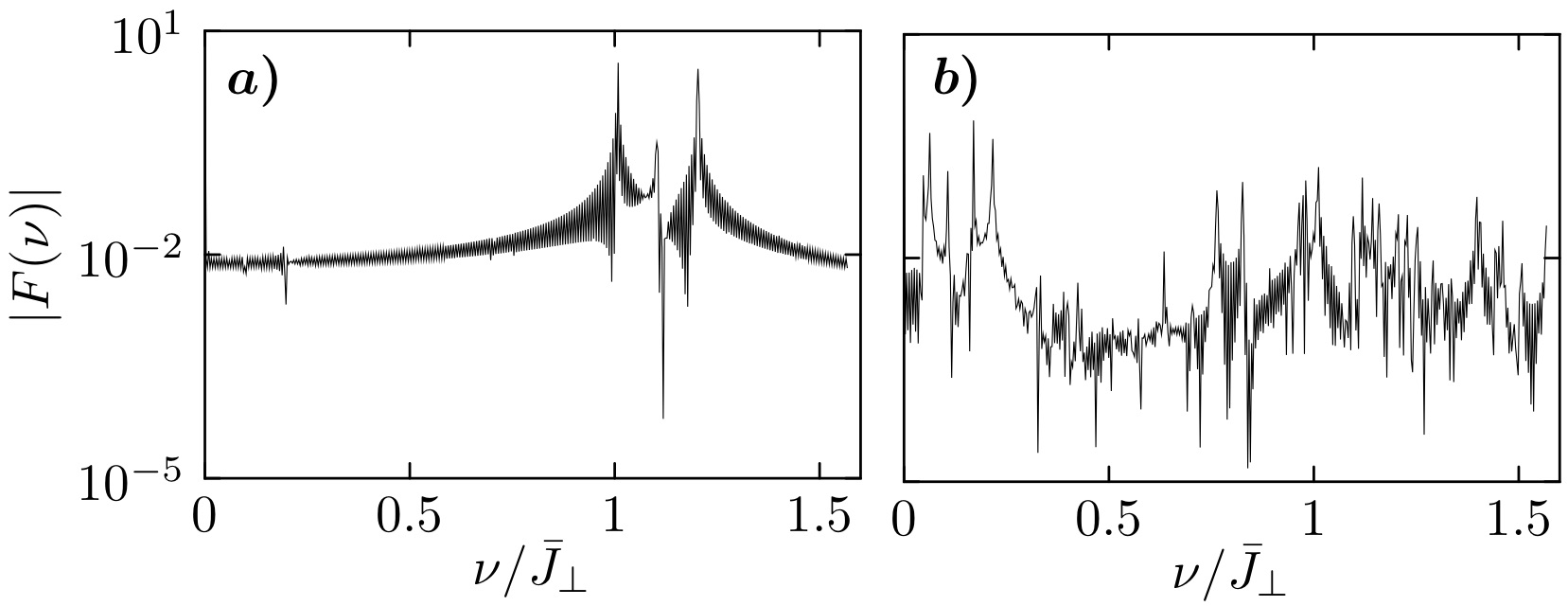}
\caption{Discrete cosine transform of $\braket{ \hat{O}(t)}$, in (a) the trivial phase at $\delta=1$, and (b) the topological phase at $\delta=0$. Inset in (b): Magnified view of the spectrum near $\nu=0$. Three large peaks, corresponding to Rabi oscillations between the ground state and the lowest three excited states, are prominently visible in (a). (b) shows five peaks near $\nu=0$, corresponding to five of six pairwise Rabi oscillations between the four lowest eigenstates of $\hat{H}(\boldsymbol{g}^{\rm opt})$. Here we chose $\epsilon = 0.05$. The peaks converge as $\epsilon \rightarrow 0$.}
\label{fig: BSsupp}
\end{figure}

As in the two-level case, the eigenvalue differences, $\xi_\alpha - \xi_\beta$, of $\hat{H}(\boldsymbol{g}^{\rm opt})$ can be easily obtained from the Fourier peaks of any generic observable $\braket{ \hat{O}(t) }$. We chose the observable to be measured as $ \hat{O} = \sum_j (-1)^{j_x+j_y} \hat\sigma^z_j$, since it exhibits oscillations with relatively large amplitudes and therefore large Fourier peaks. The heights of the Fourier peaks are determined by a few factors. First, the Fourier peak at frequency $\Delta_{\alpha\beta} = \xi_\alpha-\xi_\beta$ depends on the ratio $\Omega_{\alpha\beta}^2/(\Omega_{\alpha\beta}^2 + \Delta_{\alpha\beta}^2/4)$ and the matrix element of $\hat{O}$ between $\ket{ \xi_\alpha }$ and $\ket{ \xi_\beta }$. The second factor, which is the more important factor in our protocol, is the occupation of $\ket{\xi_\alpha}$ and $\ket{\xi_\beta}$. Our initial state is a thermal mixture, $\rho = \exp(-\tilde{H}_A) = \sum_\alpha \exp(-\xi_\alpha) \ket{\xi_\alpha}\bra{\xi_\alpha}$. The occupations decrease exponentially with $\xi_\alpha$, and therefore so do the Fourier peak heights. Consequently, for moderate $\delta$, the most prominent Fourier peaks correspond to Rabi oscillations between the ground state $\ket{\xi_0}$ and an excited state $\ket{\xi_{\alpha\neq 0}}$. As one approaches the topological phase by decreasing $\delta$, the values of $\xi_\alpha$ for the low-lying excited states decrease, thus increasing the occupation in these levels. Then, one also begins seeing prominent Fourier peaks due to Rabi oscillations between excited states, as demonstrated in Fig.~\ref{fig: BSsupp}.

Finally, we comment on the method used to convert $\braket{ O(t) }$ to frequency space and extract the peaks. While we always referred to Fourier peaks above, which are peaks in the Fourier transform of $\braket{ \hat{O}(t) }$, one can also obtain the spectrum from a discrete cosine transform (DCT) or discrete sine transform of $\braket{ \hat{O}(t) }$. These transforms involve expansions in $\cos\omega t$ or $\sin\omega t$ instead of $\exp(i\omega t)$. We find that using the DCT, the frequency resolution we can achieve for a given maximum time evolution $T_{\rm max}$ is roughly halved. This can be particularly useful when approaching the topological phase, which has degenerate levels.

Figure~\ref{fig: BSsupp}(a) plots the DCT of $\braket{ \hat{O}(t) }$, denoted $F(\nu)$, at $\delta = 1, \epsilon= 0.05J$, and Fig.~\ref{fig: BSsupp}(b) plots $F(\nu)$ at $\delta = 0, \epsilon=0.05J$. The ringing around the peaks is a common occurrence in the DCT. Nevertheless, three prominent peaks are clearly visible in Fig.~\ref{fig: BSsupp}(a), corresponding to Rabi oscillations between $\ket{\xi_0}$ and $\ket{\xi_{1,2,3}}$. The three peaks converge to the same nonzero value as $\epsilon$ is decreased, and become degenerate at $\epsilon=0$ due to the symmetry of the Heisenberg model, as can be seen in Fig.~\ref{fig: ES}(c) in the main text. The spectrum in Fig.~\ref{fig: BSsupp}(b) shows five peaks close together at small frequencies (magnified in the inset). These peaks arise from six pairwise Rabi oscillations between the four lowest eigenstates of $\hat{H}(\boldsymbol{g}^{\rm opt})$. Of the six peaks, five are resolvable. The peaks converge to $\nu=0$ as $\epsilon \rightarrow 0$, indicating the fourfold degeneracy of the eigenstates and thus the topological phase.

\bibliography{refs}

\begin{thebibliography}{60}%
\makeatletter
\providecommand \@ifxundefined [1]{%
 \@ifx{#1\undefined}
}%
\providecommand \@ifnum [1]{%
 \ifnum #1\expandafter \@firstoftwo
 \else \expandafter \@secondoftwo
 \fi
}%
\providecommand \@ifx [1]{%
 \ifx #1\expandafter \@firstoftwo
 \else \expandafter \@secondoftwo
 \fi
}%
\providecommand \natexlab [1]{#1}%
\providecommand \enquote  [1]{``#1''}%
\providecommand \bibnamefont  [1]{#1}%
\providecommand \bibfnamefont [1]{#1}%
\providecommand \citenamefont [1]{#1}%
\providecommand \href@noop [0]{\@secondoftwo}%
\providecommand \href [0]{\begingroup \@sanitize@url \@href}%
\providecommand \@href[1]{\@@startlink{#1}\@@href}%
\providecommand \@@href[1]{\endgroup#1\@@endlink}%
\providecommand \@sanitize@url [0]{\catcode `\\12\catcode `\$12\catcode
  `\&12\catcode `\#12\catcode `\^12\catcode `\_12\catcode `\%12\relax}%
\providecommand \@@startlink[1]{}%
\providecommand \@@endlink[0]{}%
\providecommand \url  [0]{\begingroup\@sanitize@url \@url }%
\providecommand \@url [1]{\endgroup\@href {#1}{\urlprefix }}%
\providecommand \urlprefix  [0]{URL }%
\providecommand \Eprint [0]{\href }%
\providecommand \doibase [0]{http://dx.doi.org/}%
\providecommand \selectlanguage [0]{\@gobble}%
\providecommand \bibinfo  [0]{\@secondoftwo}%
\providecommand \bibfield  [0]{\@secondoftwo}%
\providecommand \translation [1]{[#1]}%
\providecommand \BibitemOpen [0]{}%
\providecommand \bibitemStop [0]{}%
\providecommand \bibitemNoStop [0]{.\EOS\space}%
\providecommand \EOS [0]{\spacefactor3000\relax}%
\providecommand \BibitemShut  [1]{\csname bibitem#1\endcsname}%
\let\auto@bib@innerbib\@empty
%</preamble>
\bibitem [{\citenamefont {National Academies~of Sciences}\ and\ \citenamefont
  {Medicine}(2020)}]{NAP25613}%
  \BibitemOpen
  \bibfield  {author} {\bibinfo {author} {\bibfnamefont {E.}~\bibnamefont
  {National Academies~of Sciences}}\ and\ \bibinfo {author} {\bibnamefont
  {Medicine}},\ }\href {\doibase 10.17226/25613} {\emph {\bibinfo {title}
  {Manipulating Quantum Systems: An Assessment of Atomic, Molecular, and
  Optical Physics in the United States}}}\ (\bibinfo  {publisher} {The National
  Academies Press},\ \bibinfo {address} {Washington, DC},\ \bibinfo {year}
  {2020})\BibitemShut {NoStop}%
\bibitem [{\citenamefont {Altman}\ \emph {et~al.}(2021)\citenamefont {Altman},
  \citenamefont {Brown}, \citenamefont {Carleo}, \citenamefont {Carr},
  \citenamefont {Demler}, \citenamefont {Chin}, \citenamefont {DeMarco},
  \citenamefont {Economou}, \citenamefont {Eriksson}, \citenamefont {Fu} \emph
  {et~al.}}]{altman2021quantum}%
  \BibitemOpen
  \bibfield  {author} {\bibinfo {author} {\bibfnamefont {E.}~\bibnamefont
  {Altman}}, \bibinfo {author} {\bibfnamefont {K.~R.}\ \bibnamefont {Brown}},
  \bibinfo {author} {\bibfnamefont {G.}~\bibnamefont {Carleo}}, \bibinfo
  {author} {\bibfnamefont {L.~D.}\ \bibnamefont {Carr}}, \bibinfo {author}
  {\bibfnamefont {E.}~\bibnamefont {Demler}}, \bibinfo {author} {\bibfnamefont
  {C.}~\bibnamefont {Chin}}, \bibinfo {author} {\bibfnamefont {B.}~\bibnamefont
  {DeMarco}}, \bibinfo {author} {\bibfnamefont {S.~E.}\ \bibnamefont
  {Economou}}, \bibinfo {author} {\bibfnamefont {M.~A.}\ \bibnamefont
  {Eriksson}}, \bibinfo {author} {\bibfnamefont {K.-M.~C.}\ \bibnamefont {Fu}},
   \emph {et~al.},\ }\href {\doibase 10.1103/PRXQuantum.2.017003} {\bibfield
  {journal} {\bibinfo  {journal} {PRX Quantum}\ }\textbf {\bibinfo {volume}
  {2}},\ \bibinfo {pages} {017003} (\bibinfo {year} {2021})}\BibitemShut
  {NoStop}%
\bibitem [{\citenamefont {Gross}\ and\ \citenamefont {Bloch}(2017)}]{Gross995}%
  \BibitemOpen
  \bibfield  {author} {\bibinfo {author} {\bibfnamefont {C.}~\bibnamefont
  {Gross}}\ and\ \bibinfo {author} {\bibfnamefont {I.}~\bibnamefont {Bloch}},\
  }\href {\doibase 10.1126/science.aal3837} {\bibfield  {journal} {\bibinfo
  {journal} {Science}\ }\textbf {\bibinfo {volume} {357}},\ \bibinfo {pages}
  {995} (\bibinfo {year} {2017})}\BibitemShut {NoStop}%
\bibitem [{\citenamefont {Sun}\ \emph {et~al.}(2021)\citenamefont {Sun},
  \citenamefont {Yang}, \citenamefont {Wang}, \citenamefont {Zhou},
  \citenamefont {Su}, \citenamefont {Dai}, \citenamefont {Yuan},\ and\
  \citenamefont {Pan}}]{sun2020realization}%
  \BibitemOpen
  \bibfield  {author} {\bibinfo {author} {\bibfnamefont {H.}~\bibnamefont
  {Sun}}, \bibinfo {author} {\bibfnamefont {B.}~\bibnamefont {Yang}}, \bibinfo
  {author} {\bibfnamefont {H.-Y.}\ \bibnamefont {Wang}}, \bibinfo {author}
  {\bibfnamefont {Z.-Y.}\ \bibnamefont {Zhou}}, \bibinfo {author}
  {\bibfnamefont {G.-X.}\ \bibnamefont {Su}}, \bibinfo {author} {\bibfnamefont
  {H.-N.}\ \bibnamefont {Dai}}, \bibinfo {author} {\bibfnamefont {Z.-S.}\
  \bibnamefont {Yuan}}, \ and\ \bibinfo {author} {\bibfnamefont {J.-W.}\
  \bibnamefont {Pan}},\ }\href {\doibase 10.1038/s41567-021-01277-1} {\bibfield
   {journal} {\bibinfo  {journal} {Nature Physics}\ }\textbf {\bibinfo {volume}
  {17}},\ \bibinfo {pages} {990} (\bibinfo {year} {2021})}\BibitemShut
  {NoStop}%
\bibitem [{\citenamefont {Mazurenko}\ \emph {et~al.}(2017)\citenamefont
  {Mazurenko}, \citenamefont {Chiu}, \citenamefont {Ji}, \citenamefont
  {Parsons}, \citenamefont {Kan{\'a}sz-Nagy}, \citenamefont {Schmidt},
  \citenamefont {Grusdt}, \citenamefont {Demler}, \citenamefont {Greif},\ and\
  \citenamefont {Greiner}}]{Mazurenko2017}%
  \BibitemOpen
  \bibfield  {author} {\bibinfo {author} {\bibfnamefont {A.}~\bibnamefont
  {Mazurenko}}, \bibinfo {author} {\bibfnamefont {C.~S.}\ \bibnamefont {Chiu}},
  \bibinfo {author} {\bibfnamefont {G.}~\bibnamefont {Ji}}, \bibinfo {author}
  {\bibfnamefont {M.~F.}\ \bibnamefont {Parsons}}, \bibinfo {author}
  {\bibfnamefont {M.}~\bibnamefont {Kan{\'a}sz-Nagy}}, \bibinfo {author}
  {\bibfnamefont {R.}~\bibnamefont {Schmidt}}, \bibinfo {author} {\bibfnamefont
  {F.}~\bibnamefont {Grusdt}}, \bibinfo {author} {\bibfnamefont
  {E.}~\bibnamefont {Demler}}, \bibinfo {author} {\bibfnamefont
  {D.}~\bibnamefont {Greif}}, \ and\ \bibinfo {author} {\bibfnamefont
  {M.}~\bibnamefont {Greiner}},\ }\href {\doibase 10.1038/nature22362}
  {\bibfield  {journal} {\bibinfo  {journal} {Nature}\ }\textbf {\bibinfo
  {volume} {545}},\ \bibinfo {pages} {462} (\bibinfo {year}
  {2017})}\BibitemShut {NoStop}%
\bibitem [{\citenamefont {Hartke}\ \emph {et~al.}(2020)\citenamefont {Hartke},
  \citenamefont {Oreg}, \citenamefont {Jia},\ and\ \citenamefont
  {Zwierlein}}]{hartke2020doublon}%
  \BibitemOpen
  \bibfield  {author} {\bibinfo {author} {\bibfnamefont {T.}~\bibnamefont
  {Hartke}}, \bibinfo {author} {\bibfnamefont {B.}~\bibnamefont {Oreg}},
  \bibinfo {author} {\bibfnamefont {N.}~\bibnamefont {Jia}}, \ and\ \bibinfo
  {author} {\bibfnamefont {M.}~\bibnamefont {Zwierlein}},\ }\href {\doibase
  10.1103/PhysRevLett.125.113601} {\bibfield  {journal} {\bibinfo  {journal}
  {Phys. Rev. Lett.}\ }\textbf {\bibinfo {volume} {125}},\ \bibinfo {pages}
  {113601} (\bibinfo {year} {2020})}\BibitemShut {NoStop}%
\bibitem [{\citenamefont {Vijayan}\ \emph {et~al.}(2020)\citenamefont
  {Vijayan}, \citenamefont {Sompet}, \citenamefont {Salomon}, \citenamefont
  {Koepsell}, \citenamefont {Hirthe}, \citenamefont {Bohrdt}, \citenamefont
  {Grusdt}, \citenamefont {Bloch},\ and\ \citenamefont
  {Gross}}]{vijayan2020time}%
  \BibitemOpen
  \bibfield  {author} {\bibinfo {author} {\bibfnamefont {J.}~\bibnamefont
  {Vijayan}}, \bibinfo {author} {\bibfnamefont {P.}~\bibnamefont {Sompet}},
  \bibinfo {author} {\bibfnamefont {G.}~\bibnamefont {Salomon}}, \bibinfo
  {author} {\bibfnamefont {J.}~\bibnamefont {Koepsell}}, \bibinfo {author}
  {\bibfnamefont {S.}~\bibnamefont {Hirthe}}, \bibinfo {author} {\bibfnamefont
  {A.}~\bibnamefont {Bohrdt}}, \bibinfo {author} {\bibfnamefont
  {F.}~\bibnamefont {Grusdt}}, \bibinfo {author} {\bibfnamefont
  {I.}~\bibnamefont {Bloch}}, \ and\ \bibinfo {author} {\bibfnamefont
  {C.}~\bibnamefont {Gross}},\ }\href {\doibase 10.1126/science.aay2354}
  {\bibfield  {journal} {\bibinfo  {journal} {Science}\ }\textbf {\bibinfo
  {volume} {367}},\ \bibinfo {pages} {186} (\bibinfo {year}
  {2020})}\BibitemShut {NoStop}%
\bibitem [{\citenamefont {Holten}\ \emph {et~al.}(2021)\citenamefont {Holten},
  \citenamefont {Bayha}, \citenamefont {Subramanian}, \citenamefont {Heintze},
  \citenamefont {Preiss},\ and\ \citenamefont
  {Jochim}}]{holten2021observation}%
  \BibitemOpen
  \bibfield  {author} {\bibinfo {author} {\bibfnamefont {M.}~\bibnamefont
  {Holten}}, \bibinfo {author} {\bibfnamefont {L.}~\bibnamefont {Bayha}},
  \bibinfo {author} {\bibfnamefont {K.}~\bibnamefont {Subramanian}}, \bibinfo
  {author} {\bibfnamefont {C.}~\bibnamefont {Heintze}}, \bibinfo {author}
  {\bibfnamefont {P.~M.}\ \bibnamefont {Preiss}}, \ and\ \bibinfo {author}
  {\bibfnamefont {S.}~\bibnamefont {Jochim}},\ }\href
  {https://link.aps.org/doi/10.1103/PhysRevLett.126.020401} {\bibfield
  {journal} {\bibinfo  {journal} {Phys. Rev. Lett.}\ }\textbf {\bibinfo
  {volume} {126}},\ \bibinfo {pages} {020401} (\bibinfo {year}
  {2021})}\BibitemShut {NoStop}%
\bibitem [{\citenamefont {Nichols}\ \emph {et~al.}(2019)\citenamefont
  {Nichols}, \citenamefont {Cheuk}, \citenamefont {Okan}, \citenamefont
  {Hartke}, \citenamefont {Mendez}, \citenamefont {Senthil}, \citenamefont
  {Khatami}, \citenamefont {Zhang},\ and\ \citenamefont
  {Zwierlein}}]{Nichols383}%
  \BibitemOpen
  \bibfield  {author} {\bibinfo {author} {\bibfnamefont {M.~A.}\ \bibnamefont
  {Nichols}}, \bibinfo {author} {\bibfnamefont {L.~W.}\ \bibnamefont {Cheuk}},
  \bibinfo {author} {\bibfnamefont {M.}~\bibnamefont {Okan}}, \bibinfo {author}
  {\bibfnamefont {T.~R.}\ \bibnamefont {Hartke}}, \bibinfo {author}
  {\bibfnamefont {E.}~\bibnamefont {Mendez}}, \bibinfo {author} {\bibfnamefont
  {T.}~\bibnamefont {Senthil}}, \bibinfo {author} {\bibfnamefont
  {E.}~\bibnamefont {Khatami}}, \bibinfo {author} {\bibfnamefont
  {H.}~\bibnamefont {Zhang}}, \ and\ \bibinfo {author} {\bibfnamefont {M.~W.}\
  \bibnamefont {Zwierlein}},\ }\href {\doibase 10.1126/science.aat4387}
  {\bibfield  {journal} {\bibinfo  {journal} {Science}\ }\textbf {\bibinfo
  {volume} {363}},\ \bibinfo {pages} {383} (\bibinfo {year}
  {2019})}\BibitemShut {NoStop}%
\bibitem [{\citenamefont {Brown}\ \emph {et~al.}(2019)\citenamefont {Brown},
  \citenamefont {Mitra}, \citenamefont {Guardado-Sanchez}, \citenamefont
  {Nourafkan}, \citenamefont {Reymbaut}, \citenamefont {H{\'e}bert},
  \citenamefont {Bergeron}, \citenamefont {Tremblay}, \citenamefont {Kokalj},
  \citenamefont {Huse}, \citenamefont {Schau{\ss}},\ and\ \citenamefont
  {Bakr}}]{Brown379}%
  \BibitemOpen
  \bibfield  {author} {\bibinfo {author} {\bibfnamefont {P.~T.}\ \bibnamefont
  {Brown}}, \bibinfo {author} {\bibfnamefont {D.}~\bibnamefont {Mitra}},
  \bibinfo {author} {\bibfnamefont {E.}~\bibnamefont {Guardado-Sanchez}},
  \bibinfo {author} {\bibfnamefont {R.}~\bibnamefont {Nourafkan}}, \bibinfo
  {author} {\bibfnamefont {A.}~\bibnamefont {Reymbaut}}, \bibinfo {author}
  {\bibfnamefont {C.-D.}\ \bibnamefont {H{\'e}bert}}, \bibinfo {author}
  {\bibfnamefont {S.}~\bibnamefont {Bergeron}}, \bibinfo {author}
  {\bibfnamefont {A.-M.~S.}\ \bibnamefont {Tremblay}}, \bibinfo {author}
  {\bibfnamefont {J.}~\bibnamefont {Kokalj}}, \bibinfo {author} {\bibfnamefont
  {D.~A.}\ \bibnamefont {Huse}}, \bibinfo {author} {\bibfnamefont
  {P.}~\bibnamefont {Schau{\ss}}}, \ and\ \bibinfo {author} {\bibfnamefont
  {W.~S.}\ \bibnamefont {Bakr}},\ }\href {\doibase 10.1126/science.aat4134}
  {\bibfield  {journal} {\bibinfo  {journal} {Science}\ }\textbf {\bibinfo
  {volume} {363}},\ \bibinfo {pages} {379} (\bibinfo {year}
  {2019})}\BibitemShut {NoStop}%
\bibitem [{\citenamefont {Scholl}\ \emph {et~al.}(2021)\citenamefont {Scholl},
  \citenamefont {Schuler}, \citenamefont {Williams}, \citenamefont
  {Eberharter}, \citenamefont {Barredo}, \citenamefont {Schymik}, \citenamefont
  {Lienhard}, \citenamefont {Henry}, \citenamefont {Lang}, \citenamefont
  {Lahaye}, \citenamefont {L{\"a}uchli},\ and\ \citenamefont
  {Browaeys}}]{scholl2020programmable}%
  \BibitemOpen
  \bibfield  {author} {\bibinfo {author} {\bibfnamefont {P.}~\bibnamefont
  {Scholl}}, \bibinfo {author} {\bibfnamefont {M.}~\bibnamefont {Schuler}},
  \bibinfo {author} {\bibfnamefont {H.~J.}\ \bibnamefont {Williams}}, \bibinfo
  {author} {\bibfnamefont {A.~A.}\ \bibnamefont {Eberharter}}, \bibinfo
  {author} {\bibfnamefont {D.}~\bibnamefont {Barredo}}, \bibinfo {author}
  {\bibfnamefont {K.-N.}\ \bibnamefont {Schymik}}, \bibinfo {author}
  {\bibfnamefont {V.}~\bibnamefont {Lienhard}}, \bibinfo {author}
  {\bibfnamefont {L.-P.}\ \bibnamefont {Henry}}, \bibinfo {author}
  {\bibfnamefont {T.~C.}\ \bibnamefont {Lang}}, \bibinfo {author}
  {\bibfnamefont {T.}~\bibnamefont {Lahaye}}, \bibinfo {author} {\bibfnamefont
  {A.~M.}\ \bibnamefont {L{\"a}uchli}}, \ and\ \bibinfo {author} {\bibfnamefont
  {A.}~\bibnamefont {Browaeys}},\ }\href {\doibase 10.1038/s41586-021-03585-1}
  {\bibfield  {journal} {\bibinfo  {journal} {Nature}\ }\textbf {\bibinfo
  {volume} {595}},\ \bibinfo {pages} {233} (\bibinfo {year}
  {2021})}\BibitemShut {NoStop}%
\bibitem [{\citenamefont {Ebadi}\ \emph {et~al.}(2021)\citenamefont {Ebadi},
  \citenamefont {Wang}, \citenamefont {Levine}, \citenamefont {Keesling},
  \citenamefont {Semeghini}, \citenamefont {Omran}, \citenamefont {Bluvstein},
  \citenamefont {Samajdar}, \citenamefont {Pichler}, \citenamefont {Ho},
  \citenamefont {Choi}, \citenamefont {Sachdev}, \citenamefont {Greiner},
  \citenamefont {Vuleti{\'c}},\ and\ \citenamefont {Lukin}}]{ebadi2020quantum}%
  \BibitemOpen
  \bibfield  {author} {\bibinfo {author} {\bibfnamefont {S.}~\bibnamefont
  {Ebadi}}, \bibinfo {author} {\bibfnamefont {T.~T.}\ \bibnamefont {Wang}},
  \bibinfo {author} {\bibfnamefont {H.}~\bibnamefont {Levine}}, \bibinfo
  {author} {\bibfnamefont {A.}~\bibnamefont {Keesling}}, \bibinfo {author}
  {\bibfnamefont {G.}~\bibnamefont {Semeghini}}, \bibinfo {author}
  {\bibfnamefont {A.}~\bibnamefont {Omran}}, \bibinfo {author} {\bibfnamefont
  {D.}~\bibnamefont {Bluvstein}}, \bibinfo {author} {\bibfnamefont
  {R.}~\bibnamefont {Samajdar}}, \bibinfo {author} {\bibfnamefont
  {H.}~\bibnamefont {Pichler}}, \bibinfo {author} {\bibfnamefont {W.~W.}\
  \bibnamefont {Ho}}, \bibinfo {author} {\bibfnamefont {S.}~\bibnamefont
  {Choi}}, \bibinfo {author} {\bibfnamefont {S.}~\bibnamefont {Sachdev}},
  \bibinfo {author} {\bibfnamefont {M.}~\bibnamefont {Greiner}}, \bibinfo
  {author} {\bibfnamefont {V.}~\bibnamefont {Vuleti{\'c}}}, \ and\ \bibinfo
  {author} {\bibfnamefont {M.~D.}\ \bibnamefont {Lukin}},\ }\href {\doibase
  10.1038/s41586-021-03582-4} {\bibfield  {journal} {\bibinfo  {journal}
  {Nature}\ }\textbf {\bibinfo {volume} {595}},\ \bibinfo {pages} {227}
  (\bibinfo {year} {2021})}\BibitemShut {NoStop}%
\bibitem [{\citenamefont {Semeghini}\ \emph {et~al.}(2021)\citenamefont
  {Semeghini}, \citenamefont {Levine}, \citenamefont {Keesling}, \citenamefont
  {Ebadi}, \citenamefont {Wang}, \citenamefont {Bluvstein}, \citenamefont
  {Verresen}, \citenamefont {Pichler}, \citenamefont {Kalinowski},
  \citenamefont {Samajdar}, \citenamefont {Omran}, \citenamefont {Sachdev},
  \citenamefont {Vishwanath}, \citenamefont {Greiner}, \citenamefont
  {Vuletic},\ and\ \citenamefont {Lukin}}]{semeghini2021probing}%
  \BibitemOpen
  \bibfield  {author} {\bibinfo {author} {\bibfnamefont {G.}~\bibnamefont
  {Semeghini}}, \bibinfo {author} {\bibfnamefont {H.}~\bibnamefont {Levine}},
  \bibinfo {author} {\bibfnamefont {A.}~\bibnamefont {Keesling}}, \bibinfo
  {author} {\bibfnamefont {S.}~\bibnamefont {Ebadi}}, \bibinfo {author}
  {\bibfnamefont {T.~T.}\ \bibnamefont {Wang}}, \bibinfo {author}
  {\bibfnamefont {D.}~\bibnamefont {Bluvstein}}, \bibinfo {author}
  {\bibfnamefont {R.}~\bibnamefont {Verresen}}, \bibinfo {author}
  {\bibfnamefont {H.}~\bibnamefont {Pichler}}, \bibinfo {author} {\bibfnamefont
  {M.}~\bibnamefont {Kalinowski}}, \bibinfo {author} {\bibfnamefont
  {R.}~\bibnamefont {Samajdar}}, \bibinfo {author} {\bibfnamefont
  {A.}~\bibnamefont {Omran}}, \bibinfo {author} {\bibfnamefont
  {S.}~\bibnamefont {Sachdev}}, \bibinfo {author} {\bibfnamefont
  {A.}~\bibnamefont {Vishwanath}}, \bibinfo {author} {\bibfnamefont
  {M.}~\bibnamefont {Greiner}}, \bibinfo {author} {\bibfnamefont
  {V.}~\bibnamefont {Vuletic}}, \ and\ \bibinfo {author} {\bibfnamefont
  {M.~D.}\ \bibnamefont {Lukin}},\ }\href@noop {} {} (\bibinfo {year} {2021}),\
  \Eprint {http://arxiv.org/abs/2104.04119} {arXiv:2104.04119} \BibitemShut
  {NoStop}%
\bibitem [{\citenamefont {Monroe}\ \emph {et~al.}(2021)\citenamefont {Monroe},
  \citenamefont {Campbell}, \citenamefont {Duan}, \citenamefont {Gong},
  \citenamefont {Gorshkov}, \citenamefont {Hess}, \citenamefont {Islam},
  \citenamefont {Kim}, \citenamefont {Linke}, \citenamefont {Pagano},
  \citenamefont {Richerme}, \citenamefont {Senko},\ and\ \citenamefont
  {Yao}}]{RevModPhys.93.025001}%
  \BibitemOpen
  \bibfield  {author} {\bibinfo {author} {\bibfnamefont {C.}~\bibnamefont
  {Monroe}}, \bibinfo {author} {\bibfnamefont {W.~C.}\ \bibnamefont
  {Campbell}}, \bibinfo {author} {\bibfnamefont {L.-M.}\ \bibnamefont {Duan}},
  \bibinfo {author} {\bibfnamefont {Z.-X.}\ \bibnamefont {Gong}}, \bibinfo
  {author} {\bibfnamefont {A.~V.}\ \bibnamefont {Gorshkov}}, \bibinfo {author}
  {\bibfnamefont {P.~W.}\ \bibnamefont {Hess}}, \bibinfo {author}
  {\bibfnamefont {R.}~\bibnamefont {Islam}}, \bibinfo {author} {\bibfnamefont
  {K.}~\bibnamefont {Kim}}, \bibinfo {author} {\bibfnamefont {N.~M.}\
  \bibnamefont {Linke}}, \bibinfo {author} {\bibfnamefont {G.}~\bibnamefont
  {Pagano}}, \bibinfo {author} {\bibfnamefont {P.}~\bibnamefont {Richerme}},
  \bibinfo {author} {\bibfnamefont {C.}~\bibnamefont {Senko}}, \ and\ \bibinfo
  {author} {\bibfnamefont {N.~Y.}\ \bibnamefont {Yao}},\ }\href {\doibase
  10.1103/RevModPhys.93.025001} {\bibfield  {journal} {\bibinfo  {journal}
  {Rev. Mod. Phys.}\ }\textbf {\bibinfo {volume} {93}},\ \bibinfo {pages}
  {025001} (\bibinfo {year} {2021})}\BibitemShut {NoStop}%
\bibitem [{\citenamefont {Kokail}\ \emph {et~al.}(2019)\citenamefont {Kokail},
  \citenamefont {Maier}, \citenamefont {van Bijnen}, \citenamefont {Brydges},
  \citenamefont {Joshi}, \citenamefont {Jurcevic}, \citenamefont {Muschik},
  \citenamefont {Silvi}, \citenamefont {Blatt}, \citenamefont {Roos} \emph
  {et~al.}}]{kokail2019self}%
  \BibitemOpen
  \bibfield  {author} {\bibinfo {author} {\bibfnamefont {C.}~\bibnamefont
  {Kokail}}, \bibinfo {author} {\bibfnamefont {C.}~\bibnamefont {Maier}},
  \bibinfo {author} {\bibfnamefont {R.}~\bibnamefont {van Bijnen}}, \bibinfo
  {author} {\bibfnamefont {T.}~\bibnamefont {Brydges}}, \bibinfo {author}
  {\bibfnamefont {M.~K.}\ \bibnamefont {Joshi}}, \bibinfo {author}
  {\bibfnamefont {P.}~\bibnamefont {Jurcevic}}, \bibinfo {author}
  {\bibfnamefont {C.~A.}\ \bibnamefont {Muschik}}, \bibinfo {author}
  {\bibfnamefont {P.}~\bibnamefont {Silvi}}, \bibinfo {author} {\bibfnamefont
  {R.}~\bibnamefont {Blatt}}, \bibinfo {author} {\bibfnamefont {C.~F.}\
  \bibnamefont {Roos}},  \emph {et~al.},\ }\href {\doibase
  10.1038/s41586-019-1177-4} {\bibfield  {journal} {\bibinfo  {journal}
  {Nature}\ }\textbf {\bibinfo {volume} {569}},\ \bibinfo {pages} {355}
  (\bibinfo {year} {2019})}\BibitemShut {NoStop}%
\bibitem [{\citenamefont {Gross}\ and\ \citenamefont
  {Bakr}(2020)}]{gross2020quantum}%
  \BibitemOpen
  \bibfield  {author} {\bibinfo {author} {\bibfnamefont {C.}~\bibnamefont
  {Gross}}\ and\ \bibinfo {author} {\bibfnamefont {W.~S.}\ \bibnamefont
  {Bakr}},\ }\href@noop {} {} (\bibinfo {year} {2020}),\ \Eprint
  {http://arxiv.org/abs/2010.15407} {arXiv:2010.15407} \BibitemShut {NoStop}%
\bibitem [{\citenamefont {Lukin}\ \emph {et~al.}(2019)\citenamefont {Lukin},
  \citenamefont {Rispoli}, \citenamefont {Schittko}, \citenamefont {Tai},
  \citenamefont {Kaufman}, \citenamefont {Choi}, \citenamefont {Khemani},
  \citenamefont {L{\'e}onard},\ and\ \citenamefont {Greiner}}]{Lukin256}%
  \BibitemOpen
  \bibfield  {author} {\bibinfo {author} {\bibfnamefont {A.}~\bibnamefont
  {Lukin}}, \bibinfo {author} {\bibfnamefont {M.}~\bibnamefont {Rispoli}},
  \bibinfo {author} {\bibfnamefont {R.}~\bibnamefont {Schittko}}, \bibinfo
  {author} {\bibfnamefont {M.~E.}\ \bibnamefont {Tai}}, \bibinfo {author}
  {\bibfnamefont {A.~M.}\ \bibnamefont {Kaufman}}, \bibinfo {author}
  {\bibfnamefont {S.}~\bibnamefont {Choi}}, \bibinfo {author} {\bibfnamefont
  {V.}~\bibnamefont {Khemani}}, \bibinfo {author} {\bibfnamefont
  {J.}~\bibnamefont {L{\'e}onard}}, \ and\ \bibinfo {author} {\bibfnamefont
  {M.}~\bibnamefont {Greiner}},\ }\href {\doibase 10.1126/science.aau0818}
  {\bibfield  {journal} {\bibinfo  {journal} {Science}\ }\textbf {\bibinfo
  {volume} {364}},\ \bibinfo {pages} {256} (\bibinfo {year}
  {2019})}\BibitemShut {NoStop}%
\bibitem [{\citenamefont {Cerezo}\ \emph
  {et~al.}(2020{\natexlab{a}})\citenamefont {Cerezo}, \citenamefont
  {Arrasmith}, \citenamefont {Babbush}, \citenamefont {Benjamin}, \citenamefont
  {Endo}, \citenamefont {Fujii}, \citenamefont {McClean}, \citenamefont
  {Mitarai}, \citenamefont {Yuan}, \citenamefont {Cincio} \emph
  {et~al.}}]{cerezo2020variational}%
  \BibitemOpen
  \bibfield  {author} {\bibinfo {author} {\bibfnamefont {M.}~\bibnamefont
  {Cerezo}}, \bibinfo {author} {\bibfnamefont {A.}~\bibnamefont {Arrasmith}},
  \bibinfo {author} {\bibfnamefont {R.}~\bibnamefont {Babbush}}, \bibinfo
  {author} {\bibfnamefont {S.~C.}\ \bibnamefont {Benjamin}}, \bibinfo {author}
  {\bibfnamefont {S.}~\bibnamefont {Endo}}, \bibinfo {author} {\bibfnamefont
  {K.}~\bibnamefont {Fujii}}, \bibinfo {author} {\bibfnamefont {J.~R.}\
  \bibnamefont {McClean}}, \bibinfo {author} {\bibfnamefont {K.}~\bibnamefont
  {Mitarai}}, \bibinfo {author} {\bibfnamefont {X.}~\bibnamefont {Yuan}},
  \bibinfo {author} {\bibfnamefont {L.}~\bibnamefont {Cincio}},  \emph
  {et~al.},\ }\href@noop {} {\bibfield  {journal} {\bibinfo  {journal} {arXiv
  preprint arXiv:2012.09265}\ } (\bibinfo {year}
  {2020}{\natexlab{a}})}\BibitemShut {NoStop}%
\bibitem [{\citenamefont {Bairey}\ \emph {et~al.}(2019)\citenamefont {Bairey},
  \citenamefont {Arad},\ and\ \citenamefont {Lindner}}]{bairey2019learning}%
  \BibitemOpen
  \bibfield  {author} {\bibinfo {author} {\bibfnamefont {E.}~\bibnamefont
  {Bairey}}, \bibinfo {author} {\bibfnamefont {I.}~\bibnamefont {Arad}}, \ and\
  \bibinfo {author} {\bibfnamefont {N.~H.}\ \bibnamefont {Lindner}},\ }\href
  {\doibase 10.1103/physrevlett.122.020504} {\bibfield  {journal} {\bibinfo
  {journal} {Phys. Rev. Lett.}\ }\textbf {\bibinfo {volume} {122}},\ \bibinfo
  {pages} {020504} (\bibinfo {year} {2019})}\BibitemShut {NoStop}%
\bibitem [{\citenamefont {Qi}\ and\ \citenamefont
  {Ranard}(2019)}]{qi2019determining}%
  \BibitemOpen
  \bibfield  {author} {\bibinfo {author} {\bibfnamefont {X.-L.}\ \bibnamefont
  {Qi}}\ and\ \bibinfo {author} {\bibfnamefont {D.}~\bibnamefont {Ranard}},\
  }\href {\doibase 10.22331/q-2019-07-08-159} {\bibfield  {journal} {\bibinfo
  {journal} {Quantum}\ }\textbf {\bibinfo {volume} {3}},\ \bibinfo {pages}
  {159} (\bibinfo {year} {2019})}\BibitemShut {NoStop}%
\bibitem [{\citenamefont {Li}\ \emph {et~al.}(2020)\citenamefont {Li},
  \citenamefont {Zou},\ and\ \citenamefont {Hsieh}}]{li2020hamiltonian}%
  \BibitemOpen
  \bibfield  {author} {\bibinfo {author} {\bibfnamefont {Z.}~\bibnamefont
  {Li}}, \bibinfo {author} {\bibfnamefont {L.}~\bibnamefont {Zou}}, \ and\
  \bibinfo {author} {\bibfnamefont {T.~H.}\ \bibnamefont {Hsieh}},\ }\href
  {\doibase 10.1103/physrevlett.124.160502} {\bibfield  {journal} {\bibinfo
  {journal} {Phys. Rev. Lett.}\ }\textbf {\bibinfo {volume} {124}},\ \bibinfo
  {pages} {160502} (\bibinfo {year} {2020})}\BibitemShut {NoStop}%
\bibitem [{\citenamefont {Evans}\ \emph {et~al.}(2019)\citenamefont {Evans},
  \citenamefont {Harper},\ and\ \citenamefont {Flammia}}]{evans2019scalable}%
  \BibitemOpen
  \bibfield  {author} {\bibinfo {author} {\bibfnamefont {T.~J.}\ \bibnamefont
  {Evans}}, \bibinfo {author} {\bibfnamefont {R.}~\bibnamefont {Harper}}, \
  and\ \bibinfo {author} {\bibfnamefont {S.~T.}\ \bibnamefont {Flammia}},\
  }\href@noop {} {} (\bibinfo {year} {2019}),\ \Eprint
  {http://arxiv.org/abs/1912.07636} {arXiv:1912.07636} \BibitemShut {NoStop}%
\bibitem [{\citenamefont {Kokail}\ \emph {et~al.}(2021)\citenamefont {Kokail},
  \citenamefont {van Bijnen}, \citenamefont {Elben}, \citenamefont
  {Vermersch},\ and\ \citenamefont {Zoller}}]{kokail2020entanglement}%
  \BibitemOpen
  \bibfield  {author} {\bibinfo {author} {\bibfnamefont {C.}~\bibnamefont
  {Kokail}}, \bibinfo {author} {\bibfnamefont {R.}~\bibnamefont {van Bijnen}},
  \bibinfo {author} {\bibfnamefont {A.}~\bibnamefont {Elben}}, \bibinfo
  {author} {\bibfnamefont {B.}~\bibnamefont {Vermersch}}, \ and\ \bibinfo
  {author} {\bibfnamefont {P.}~\bibnamefont {Zoller}},\ }\href {\doibase
  10.1038/s41567-021-01260-w} {\bibfield  {journal} {\bibinfo  {journal}
  {Nature Physics}\ }\textbf {\bibinfo {volume} {17}},\ \bibinfo {pages} {936}
  (\bibinfo {year} {2021})}\BibitemShut {NoStop}%
\bibitem [{\citenamefont {Choo}\ \emph {et~al.}(2018)\citenamefont {Choo},
  \citenamefont {Von~Keyserlingk}, \citenamefont {Regnault},\ and\
  \citenamefont {Neupert}}]{choo2018measurement}%
  \BibitemOpen
  \bibfield  {author} {\bibinfo {author} {\bibfnamefont {K.}~\bibnamefont
  {Choo}}, \bibinfo {author} {\bibfnamefont {C.~W.}\ \bibnamefont
  {Von~Keyserlingk}}, \bibinfo {author} {\bibfnamefont {N.}~\bibnamefont
  {Regnault}}, \ and\ \bibinfo {author} {\bibfnamefont {T.}~\bibnamefont
  {Neupert}},\ }\href {\doibase 10.1103/physrevlett.121.086808} {\bibfield
  {journal} {\bibinfo  {journal} {Phys. Rev. Lett.}\ }\textbf {\bibinfo
  {volume} {121}},\ \bibinfo {pages} {086808} (\bibinfo {year}
  {2018})}\BibitemShut {NoStop}%
\bibitem [{\citenamefont {Peruzzo}\ \emph {et~al.}(2014)\citenamefont
  {Peruzzo}, \citenamefont {McClean}, \citenamefont {Shadbolt}, \citenamefont
  {Yung}, \citenamefont {Zhou}, \citenamefont {Love}, \citenamefont
  {Aspuru-Guzik},\ and\ \citenamefont {O'Brien}}]{peruzzo2014variational}%
  \BibitemOpen
  \bibfield  {author} {\bibinfo {author} {\bibfnamefont {A.}~\bibnamefont
  {Peruzzo}}, \bibinfo {author} {\bibfnamefont {J.}~\bibnamefont {McClean}},
  \bibinfo {author} {\bibfnamefont {P.}~\bibnamefont {Shadbolt}}, \bibinfo
  {author} {\bibfnamefont {M.-H.}\ \bibnamefont {Yung}}, \bibinfo {author}
  {\bibfnamefont {X.-Q.}\ \bibnamefont {Zhou}}, \bibinfo {author}
  {\bibfnamefont {P.~J.}\ \bibnamefont {Love}}, \bibinfo {author}
  {\bibfnamefont {A.}~\bibnamefont {Aspuru-Guzik}}, \ and\ \bibinfo {author}
  {\bibfnamefont {J.~L.}\ \bibnamefont {O'Brien}},\ }\href {\doibase
  10.1038/ncomms5213} {\bibfield  {journal} {\bibinfo  {journal} {Nature
  Communications}\ }\textbf {\bibinfo {volume} {5}},\ \bibinfo {pages} {4213}
  (\bibinfo {year} {2014})}\BibitemShut {NoStop}%
\bibitem [{\citenamefont {LaRose}\ \emph {et~al.}(2019)\citenamefont {LaRose},
  \citenamefont {Tikku}, \citenamefont {O'Neel-Judy}, \citenamefont {Cincio},\
  and\ \citenamefont {Coles}}]{larose2019variational}%
  \BibitemOpen
  \bibfield  {author} {\bibinfo {author} {\bibfnamefont {R.}~\bibnamefont
  {LaRose}}, \bibinfo {author} {\bibfnamefont {A.}~\bibnamefont {Tikku}},
  \bibinfo {author} {\bibfnamefont {{\'E}.}~\bibnamefont {O'Neel-Judy}},
  \bibinfo {author} {\bibfnamefont {L.}~\bibnamefont {Cincio}}, \ and\ \bibinfo
  {author} {\bibfnamefont {P.~J.}\ \bibnamefont {Coles}},\ }\href {\doibase
  10.1038/s41534-019-0167-6} {\bibfield  {journal} {\bibinfo  {journal} {npj
  Quantum Information}\ }\textbf {\bibinfo {volume} {5}},\ \bibinfo {pages}
  {57} (\bibinfo {year} {2019})}\BibitemShut {NoStop}%
\bibitem [{\citenamefont {Johri}\ \emph {et~al.}(2017)\citenamefont {Johri},
  \citenamefont {Steiger},\ and\ \citenamefont
  {Troyer}}]{johri2017entanglement}%
  \BibitemOpen
  \bibfield  {author} {\bibinfo {author} {\bibfnamefont {S.}~\bibnamefont
  {Johri}}, \bibinfo {author} {\bibfnamefont {D.~S.}\ \bibnamefont {Steiger}},
  \ and\ \bibinfo {author} {\bibfnamefont {M.}~\bibnamefont {Troyer}},\
  }\href@noop {} {\bibfield  {journal} {\bibinfo  {journal} {Phys. Rev. B}\
  }\textbf {\bibinfo {volume} {96}},\ \bibinfo {pages} {195136} (\bibinfo
  {year} {2017})}\BibitemShut {NoStop}%
\bibitem [{\citenamefont {Suba{\c{s}}{\i}}\ \emph {et~al.}(2019)\citenamefont
  {Suba{\c{s}}{\i}}, \citenamefont {Cincio},\ and\ \citenamefont
  {Coles}}]{subacsi2019entanglement}%
  \BibitemOpen
  \bibfield  {author} {\bibinfo {author} {\bibfnamefont {Y.}~\bibnamefont
  {Suba{\c{s}}{\i}}}, \bibinfo {author} {\bibfnamefont {L.}~\bibnamefont
  {Cincio}}, \ and\ \bibinfo {author} {\bibfnamefont {P.~J.}\ \bibnamefont
  {Coles}},\ }\href@noop {} {\bibfield  {journal} {\bibinfo  {journal} {J.
  Phys. A Math. Theor.}\ }\textbf {\bibinfo {volume} {52}},\ \bibinfo {pages}
  {044001} (\bibinfo {year} {2019})}\BibitemShut {NoStop}%
\bibitem [{\citenamefont {Cerezo}\ \emph
  {et~al.}(2020{\natexlab{b}})\citenamefont {Cerezo}, \citenamefont {Sharma},
  \citenamefont {Arrasmith},\ and\ \citenamefont
  {Coles}}]{cerezo2020variational2}%
  \BibitemOpen
  \bibfield  {author} {\bibinfo {author} {\bibfnamefont {M.}~\bibnamefont
  {Cerezo}}, \bibinfo {author} {\bibfnamefont {K.}~\bibnamefont {Sharma}},
  \bibinfo {author} {\bibfnamefont {A.}~\bibnamefont {Arrasmith}}, \ and\
  \bibinfo {author} {\bibfnamefont {P.~J.}\ \bibnamefont {Coles}},\ }\href@noop
  {} {\bibfield  {journal} {\bibinfo  {journal} {arXiv preprint
  arXiv:2004.01372}\ } (\bibinfo {year} {2020}{\natexlab{b}})}\BibitemShut
  {NoStop}%
\bibitem [{\citenamefont {Lloyd}\ \emph {et~al.}(2014)\citenamefont {Lloyd},
  \citenamefont {Mohseni},\ and\ \citenamefont
  {Rebentrost}}]{lloyd2014quantum}%
  \BibitemOpen
  \bibfield  {author} {\bibinfo {author} {\bibfnamefont {S.}~\bibnamefont
  {Lloyd}}, \bibinfo {author} {\bibfnamefont {M.}~\bibnamefont {Mohseni}}, \
  and\ \bibinfo {author} {\bibfnamefont {P.}~\bibnamefont {Rebentrost}},\
  }\href@noop {} {\bibfield  {journal} {\bibinfo  {journal} {Nat. Phys.}\
  }\textbf {\bibinfo {volume} {10}},\ \bibinfo {pages} {631} (\bibinfo {year}
  {2014})}\BibitemShut {NoStop}%
\bibitem [{\citenamefont {Bravo-Prieto}\ \emph {et~al.}(2020)\citenamefont
  {Bravo-Prieto}, \citenamefont {Garc{\'\i}a-Mart{\'\i}n},\ and\ \citenamefont
  {Latorre}}]{bravo2020quantum}%
  \BibitemOpen
  \bibfield  {author} {\bibinfo {author} {\bibfnamefont {C.}~\bibnamefont
  {Bravo-Prieto}}, \bibinfo {author} {\bibfnamefont {D.}~\bibnamefont
  {Garc{\'\i}a-Mart{\'\i}n}}, \ and\ \bibinfo {author} {\bibfnamefont {J.~I.}\
  \bibnamefont {Latorre}},\ }\href@noop {} {\bibfield  {journal} {\bibinfo
  {journal} {Phys. Rev. A}\ }\textbf {\bibinfo {volume} {101}},\ \bibinfo
  {pages} {062310} (\bibinfo {year} {2020})}\BibitemShut {NoStop}%
\bibitem [{\citenamefont {Regnault}(2015)}]{regnault2015entanglement}%
  \BibitemOpen
  \bibfield  {author} {\bibinfo {author} {\bibfnamefont {N.}~\bibnamefont
  {Regnault}},\ }\href@noop {} {} (\bibinfo {year} {2015}),\ \Eprint
  {http://arxiv.org/abs/1510.07670} {arXiv:1510.07670} \BibitemShut {NoStop}%
\bibitem [{\citenamefont {Haag}(2012)}]{haag2012local}%
  \BibitemOpen
  \bibfield  {author} {\bibinfo {author} {\bibfnamefont {R.}~\bibnamefont
  {Haag}},\ }\href@noop {} {\emph {\bibinfo {title} {Local quantum physics:
  Fields, particles, algebras}}}\ (\bibinfo  {publisher} {Springer Science \&
  Business Media},\ \bibinfo {address} {Berlin},\ \bibinfo {year}
  {2012})\BibitemShut {NoStop}%
\bibitem [{\citenamefont {Casini}\ \emph {et~al.}(2011)\citenamefont {Casini},
  \citenamefont {Huerta},\ and\ \citenamefont {Myers}}]{Casini:2011aa}%
  \BibitemOpen
  \bibfield  {author} {\bibinfo {author} {\bibfnamefont {H.}~\bibnamefont
  {Casini}}, \bibinfo {author} {\bibfnamefont {M.}~\bibnamefont {Huerta}}, \
  and\ \bibinfo {author} {\bibfnamefont {R.~C.}\ \bibnamefont {Myers}},\ }\href
  {https://arxiv.org/abs/1102.0440} {\bibfield  {journal} {\bibinfo  {journal}
  {JHEP}\ }\textbf {\bibinfo {volume} {1105}},\ \bibinfo {pages} {036}
  (\bibinfo {year} {2011})}\BibitemShut {NoStop}%
\bibitem [{\citenamefont {Bisognano}\ and\ \citenamefont
  {Wichmann}(1975)}]{bisognano1975duality}%
  \BibitemOpen
  \bibfield  {author} {\bibinfo {author} {\bibfnamefont {J.~J.}\ \bibnamefont
  {Bisognano}}\ and\ \bibinfo {author} {\bibfnamefont {E.~H.}\ \bibnamefont
  {Wichmann}},\ }\href {\doibase 10.1063/1.522605} {\bibfield  {journal}
  {\bibinfo  {journal} {J. Math. Phys.}\ }\textbf {\bibinfo {volume} {16}},\
  \bibinfo {pages} {985} (\bibinfo {year} {1975})}\BibitemShut {NoStop}%
\bibitem [{Note1()}]{Note1}%
  \BibitemOpen
  \bibinfo {note} {For a generalization within CFT to finite subsystems of
  radius $R$, see Ref.~\cite {hislop1982,Casini:2011aa,Cardy_2016} who proved
  that the deformation takes a parabolic shape, $\beta ({\protect \mathbf x})=
  2\pi (R^2 - \protect \mathbf {x}^2)/(2R)$, or similarly with the chord
  length.}\BibitemShut {Stop}%
\bibitem [{\citenamefont {Pourjafarabadi}\ \emph {et~al.}(2021)\citenamefont
  {Pourjafarabadi}, \citenamefont {Najafzadeh}, \citenamefont {Vaezi},\ and\
  \citenamefont {Vaezi}}]{pourjafarabadi2021entanglement}%
  \BibitemOpen
  \bibfield  {author} {\bibinfo {author} {\bibfnamefont {M.}~\bibnamefont
  {Pourjafarabadi}}, \bibinfo {author} {\bibfnamefont {H.}~\bibnamefont
  {Najafzadeh}}, \bibinfo {author} {\bibfnamefont {M.-S.}\ \bibnamefont
  {Vaezi}}, \ and\ \bibinfo {author} {\bibfnamefont {A.}~\bibnamefont
  {Vaezi}},\ }\href@noop {} {\bibfield  {journal} {\bibinfo  {journal} {Phys.
  Rev. Res.}\ }\textbf {\bibinfo {volume} {3}},\ \bibinfo {pages} {013217}
  (\bibinfo {year} {2021})}\BibitemShut {NoStop}%
\bibitem [{\citenamefont {Eisler}\ \emph {et~al.}(2020)\citenamefont {Eisler},
  \citenamefont {Giulio}, \citenamefont {Tonni},\ and\ \citenamefont
  {Peschel}}]{eisler2020entanglement}%
  \BibitemOpen
  \bibfield  {author} {\bibinfo {author} {\bibfnamefont {V.}~\bibnamefont
  {Eisler}}, \bibinfo {author} {\bibfnamefont {G.~D.}\ \bibnamefont {Giulio}},
  \bibinfo {author} {\bibfnamefont {E.}~\bibnamefont {Tonni}}, \ and\ \bibinfo
  {author} {\bibfnamefont {I.}~\bibnamefont {Peschel}},\ }\href {\doibase
  10.1088/1742-5468/abb4da} {\bibfield  {journal} {\bibinfo  {journal} {Journal
  of Statistical Mechanics: Theory and Experiment}\ }\textbf {\bibinfo {volume}
  {2020}},\ \bibinfo {pages} {103102} (\bibinfo {year} {2020})}\BibitemShut
  {NoStop}%
\bibitem [{\citenamefont {Giudici}\ \emph {et~al.}(2018)\citenamefont
  {Giudici}, \citenamefont {Mendes-Santos}, \citenamefont {Calabrese},\ and\
  \citenamefont {Dalmonte}}]{giudici2018entanglement}%
  \BibitemOpen
  \bibfield  {author} {\bibinfo {author} {\bibfnamefont {G.}~\bibnamefont
  {Giudici}}, \bibinfo {author} {\bibfnamefont {T.}~\bibnamefont
  {Mendes-Santos}}, \bibinfo {author} {\bibfnamefont {P.}~\bibnamefont
  {Calabrese}}, \ and\ \bibinfo {author} {\bibfnamefont {M.}~\bibnamefont
  {Dalmonte}},\ }\href {\doibase 10.1103/physrevb.98.134403} {\bibfield
  {journal} {\bibinfo  {journal} {Phys. Rev. B}\ }\textbf {\bibinfo {volume}
  {98}},\ \bibinfo {pages} {134403} (\bibinfo {year} {2018})}\BibitemShut
  {NoStop}%
\bibitem [{\citenamefont {Parisen~Toldin}\ and\ \citenamefont
  {Assaad}(2018)}]{Toldin:2018aa}%
  \BibitemOpen
  \bibfield  {author} {\bibinfo {author} {\bibfnamefont {F.}~\bibnamefont
  {Parisen~Toldin}}\ and\ \bibinfo {author} {\bibfnamefont {F.~F.}\
  \bibnamefont {Assaad}},\ }\href {\doibase 10.1103/PhysRevLett.121.200602}
  {\bibfield  {journal} {\bibinfo  {journal} {Phys. Rev. Lett.}\ }\textbf
  {\bibinfo {volume} {121}},\ \bibinfo {pages} {200602} (\bibinfo {year}
  {2018})}\BibitemShut {NoStop}%
\bibitem [{\citenamefont {Dalmonte}\ \emph {et~al.}(2018)\citenamefont
  {Dalmonte}, \citenamefont {Vermersch},\ and\ \citenamefont
  {Zoller}}]{dalmonte2018quantum}%
  \BibitemOpen
  \bibfield  {author} {\bibinfo {author} {\bibfnamefont {M.}~\bibnamefont
  {Dalmonte}}, \bibinfo {author} {\bibfnamefont {B.}~\bibnamefont {Vermersch}},
  \ and\ \bibinfo {author} {\bibfnamefont {P.}~\bibnamefont {Zoller}},\ }\href
  {\doibase 10.1038/s41567-018-0151-7} {\bibfield  {journal} {\bibinfo
  {journal} {Nat. Phys.}\ }\textbf {\bibinfo {volume} {14}},\ \bibinfo {pages}
  {827} (\bibinfo {year} {2018})}\BibitemShut {NoStop}%
\bibitem [{\citenamefont {Qiu}\ \emph {et~al.}(2020)\citenamefont {Qiu},
  \citenamefont {Zou}, \citenamefont {Qi},\ and\ \citenamefont
  {Li}}]{qiu2020precise}%
  \BibitemOpen
  \bibfield  {author} {\bibinfo {author} {\bibfnamefont {X.}~\bibnamefont
  {Qiu}}, \bibinfo {author} {\bibfnamefont {J.}~\bibnamefont {Zou}}, \bibinfo
  {author} {\bibfnamefont {X.}~\bibnamefont {Qi}}, \ and\ \bibinfo {author}
  {\bibfnamefont {X.}~\bibnamefont {Li}},\ }\href {\doibase
  10.1038/s41534-020-00315-9} {\bibfield  {journal} {\bibinfo  {journal} {npj
  Quantum Information}\ }\textbf {\bibinfo {volume} {6}},\ \bibinfo {pages}
  {87} (\bibinfo {year} {2020})}\BibitemShut {NoStop}%
\bibitem [{\citenamefont {Goldman}\ \emph {et~al.}(2016)\citenamefont
  {Goldman}, \citenamefont {Budich},\ and\ \citenamefont
  {Zoller}}]{goldman2016topological}%
  \BibitemOpen
  \bibfield  {author} {\bibinfo {author} {\bibfnamefont {N.}~\bibnamefont
  {Goldman}}, \bibinfo {author} {\bibfnamefont {J.~C.}\ \bibnamefont {Budich}},
  \ and\ \bibinfo {author} {\bibfnamefont {P.}~\bibnamefont {Zoller}},\ }\href
  {\doibase 10.1038/nphys3803} {\bibfield  {journal} {\bibinfo  {journal} {Nat.
  Phys.}\ }\textbf {\bibinfo {volume} {12}},\ \bibinfo {pages} {639} (\bibinfo
  {year} {2016})}\BibitemShut {NoStop}%
\bibitem [{Note2()}]{Note2}%
  \BibitemOpen
  \bibinfo {note} {\label {footnote:g_constraint}We note that here the
  variational search is constrained to $g_j > 0$ for all $j$. In the case that
  $g_j < 0$ are allowed, a solution with $g_j = -g_{j+1}$ does exits which
  freezes the particles on the individual lattice sites. To exclude this
  solution, additional observables like nearest-neighbour tunneling amplitudes
  are required.}\BibitemShut {Stop}%
\bibitem [{\citenamefont {Schweizer}\ \emph {et~al.}(2016)\citenamefont
  {Schweizer}, \citenamefont {Lohse}, \citenamefont {Citro},\ and\
  \citenamefont {Bloch}}]{PhysRevLett.117.170405}%
  \BibitemOpen
  \bibfield  {author} {\bibinfo {author} {\bibfnamefont {C.}~\bibnamefont
  {Schweizer}}, \bibinfo {author} {\bibfnamefont {M.}~\bibnamefont {Lohse}},
  \bibinfo {author} {\bibfnamefont {R.}~\bibnamefont {Citro}}, \ and\ \bibinfo
  {author} {\bibfnamefont {I.}~\bibnamefont {Bloch}},\ }\href {\doibase
  10.1103/PhysRevLett.117.170405} {\bibfield  {journal} {\bibinfo  {journal}
  {Phys. Rev. Lett.}\ }\textbf {\bibinfo {volume} {117}},\ \bibinfo {pages}
  {170405} (\bibinfo {year} {2016})}\BibitemShut {NoStop}%
\bibitem [{\citenamefont {Ke\ss{}ler}\ and\ \citenamefont
  {Marquardt}(2014)}]{PhysRevA.89.061601}%
  \BibitemOpen
  \bibfield  {author} {\bibinfo {author} {\bibfnamefont {S.}~\bibnamefont
  {Ke\ss{}ler}}\ and\ \bibinfo {author} {\bibfnamefont {F.}~\bibnamefont
  {Marquardt}},\ }\href {\doibase 10.1103/PhysRevA.89.061601} {\bibfield
  {journal} {\bibinfo  {journal} {Phys. Rev. A}\ }\textbf {\bibinfo {volume}
  {89}},\ \bibinfo {pages} {061601} (\bibinfo {year} {2014})}\BibitemShut
  {NoStop}%
\bibitem [{\citenamefont {L\"auchli}\ and\ \citenamefont
  {Schliemann}(2012)}]{PhysRevB.85.054403}%
  \BibitemOpen
  \bibfield  {author} {\bibinfo {author} {\bibfnamefont {A.~M.}\ \bibnamefont
  {L\"auchli}}\ and\ \bibinfo {author} {\bibfnamefont {J.}~\bibnamefont
  {Schliemann}},\ }\href {\doibase 10.1103/PhysRevB.85.054403} {\bibfield
  {journal} {\bibinfo  {journal} {Phys. Rev. B}\ }\textbf {\bibinfo {volume}
  {85}},\ \bibinfo {pages} {054403} (\bibinfo {year} {2012})}\BibitemShut
  {NoStop}%
\bibitem [{Note3()}]{Note3}%
  \BibitemOpen
  \bibinfo {note} {To be precise, we consider a family of reconstructed states
  $\protect \mathaccentV {hat}05E{\rho }^\protect \text {var}_A(\protect
  \boldsymbol {g}, \beta ) = Z^{-1}_A(\beta ) \protect \tmspace +\thinmuskip
  {.1667em} e^{-\beta \protect \mathaccentV {hat}05E{H}^\protect \text
  {var}_A(\protect \boldsymbol {g})}$ with $Z_A(\beta ) = \protect \text {Tr}
  \left [e^{-\beta \protect \mathaccentV {hat}05E{H}^\protect \text
  {var}_A(\protect \boldsymbol {g})}\right ]$, and define the fidelity as
  $\protect \mathcal {F}(\protect \mathaccentV {hat}05E{\rho }^\protect \text
  {var}_A(\protect \boldsymbol {g}), \protect \mathaccentV {hat}05E{\rho }_A) =
  \protect \qopname \relax m{max}_\beta \left [ \protect \text {Tr} (\protect
  \sqrt {\protect \sqrt {\protect \mathaccentV {hat}05E{\rho }_A} \protect
  \mathaccentV {hat}05E{\rho }^{\protect \text {var}}_A(\protect \boldsymbol
  {g},\beta ) \protect \sqrt {\protect \mathaccentV {hat}05E{\rho }_A}}) \right
  ]^2$.}\BibitemShut {Stop}%
\bibitem [{\citenamefont {Pichler}\ \emph {et~al.}(2016)\citenamefont
  {Pichler}, \citenamefont {Zhu}, \citenamefont {Seif}, \citenamefont
  {Zoller},\ and\ \citenamefont {Hafezi}}]{pichler2016measurement}%
  \BibitemOpen
  \bibfield  {author} {\bibinfo {author} {\bibfnamefont {H.}~\bibnamefont
  {Pichler}}, \bibinfo {author} {\bibfnamefont {G.}~\bibnamefont {Zhu}},
  \bibinfo {author} {\bibfnamefont {A.}~\bibnamefont {Seif}}, \bibinfo {author}
  {\bibfnamefont {P.}~\bibnamefont {Zoller}}, \ and\ \bibinfo {author}
  {\bibfnamefont {M.}~\bibnamefont {Hafezi}},\ }\href
  {http://dx.doi.org/10.1103/PhysRevX.6.041033} {\bibfield  {journal} {\bibinfo
   {journal} {Phys. Rev. X}\ }\textbf {\bibinfo {volume} {6}},\ \bibinfo
  {pages} {041033} (\bibinfo {year} {2016})}\BibitemShut {NoStop}%
\bibitem [{\citenamefont {Sompet}\ \emph {et~al.}(2021)\citenamefont {Sompet},
  \citenamefont {Hirthe}, \citenamefont {Bourgund}, \citenamefont {Chalopin},
  \citenamefont {Bibo}, \citenamefont {Koepsell}, \citenamefont {Bojovi{\'c}},
  \citenamefont {Verresen}, \citenamefont {Pollmann}, \citenamefont {Salomon}
  \emph {et~al.}}]{sompet2021realising}%
  \BibitemOpen
  \bibfield  {author} {\bibinfo {author} {\bibfnamefont {P.}~\bibnamefont
  {Sompet}}, \bibinfo {author} {\bibfnamefont {S.}~\bibnamefont {Hirthe}},
  \bibinfo {author} {\bibfnamefont {D.}~\bibnamefont {Bourgund}}, \bibinfo
  {author} {\bibfnamefont {T.}~\bibnamefont {Chalopin}}, \bibinfo {author}
  {\bibfnamefont {J.}~\bibnamefont {Bibo}}, \bibinfo {author} {\bibfnamefont
  {J.}~\bibnamefont {Koepsell}}, \bibinfo {author} {\bibfnamefont
  {P.}~\bibnamefont {Bojovi{\'c}}}, \bibinfo {author} {\bibfnamefont
  {R.}~\bibnamefont {Verresen}}, \bibinfo {author} {\bibfnamefont
  {F.}~\bibnamefont {Pollmann}}, \bibinfo {author} {\bibfnamefont
  {G.}~\bibnamefont {Salomon}},  \emph {et~al.},\ }\href@noop {} {} (\bibinfo
  {year} {2021}),\ \Eprint {http://arxiv.org/abs/2103.10421} {arXiv:2103.10421}
  \BibitemShut {NoStop}%
\bibitem [{\citenamefont {Li}\ and\ \citenamefont
  {Haldane}(2008)}]{li2008entanglement}%
  \BibitemOpen
  \bibfield  {author} {\bibinfo {author} {\bibfnamefont {H.}~\bibnamefont
  {Li}}\ and\ \bibinfo {author} {\bibfnamefont {F.~D.~M.}\ \bibnamefont
  {Haldane}},\ }\href@noop {} {\bibfield  {journal} {\bibinfo  {journal} {Phys.
  Rev. Lett.}\ }\textbf {\bibinfo {volume} {101}},\ \bibinfo {pages} {010504}
  (\bibinfo {year} {2008})}\BibitemShut {NoStop}%
\bibitem [{\citenamefont {Van~Nieuwenburg}\ \emph {et~al.}(2017)\citenamefont
  {Van~Nieuwenburg}, \citenamefont {Liu},\ and\ \citenamefont
  {Huber}}]{van2017learning}%
  \BibitemOpen
  \bibfield  {author} {\bibinfo {author} {\bibfnamefont {E.~P.}\ \bibnamefont
  {Van~Nieuwenburg}}, \bibinfo {author} {\bibfnamefont {Y.-H.}\ \bibnamefont
  {Liu}}, \ and\ \bibinfo {author} {\bibfnamefont {S.~D.}\ \bibnamefont
  {Huber}},\ }\href {\doibase 10.1038/nphys4037} {\bibfield  {journal}
  {\bibinfo  {journal} {Nat. Phys.}\ }\textbf {\bibinfo {volume} {13}},\
  \bibinfo {pages} {435} (\bibinfo {year} {2017})}\BibitemShut {NoStop}%
\bibitem [{\citenamefont {Cerezo}\ \emph {et~al.}(2021)\citenamefont {Cerezo},
  \citenamefont {Sone}, \citenamefont {Volkoff}, \citenamefont {Cincio},\ and\
  \citenamefont {Coles}}]{barren_plateau}%
  \BibitemOpen
  \bibfield  {author} {\bibinfo {author} {\bibfnamefont {M.}~\bibnamefont
  {Cerezo}}, \bibinfo {author} {\bibfnamefont {A.}~\bibnamefont {Sone}},
  \bibinfo {author} {\bibfnamefont {T.}~\bibnamefont {Volkoff}}, \bibinfo
  {author} {\bibfnamefont {L.}~\bibnamefont {Cincio}}, \ and\ \bibinfo {author}
  {\bibfnamefont {P.~J.}\ \bibnamefont {Coles}},\ }\href {\doibase
  10.1038/s41467-021-21728-w} {\bibfield  {journal} {\bibinfo  {journal}
  {Nature Communications}\ }\textbf {\bibinfo {volume} {12}},\ \bibinfo {pages}
  {1791} (\bibinfo {year} {2021})}\BibitemShut {NoStop}%
\bibitem [{\citenamefont {Vidal}(2003)}]{PhysRevLett.91.147902}%
  \BibitemOpen
  \bibfield  {author} {\bibinfo {author} {\bibfnamefont {G.}~\bibnamefont
  {Vidal}},\ }\href {\doibase 10.1103/PhysRevLett.91.147902} {\bibfield
  {journal} {\bibinfo  {journal} {Phys. Rev. Lett.}\ }\textbf {\bibinfo
  {volume} {91}},\ \bibinfo {pages} {147902} (\bibinfo {year}
  {2003})}\BibitemShut {NoStop}%
\bibitem [{\citenamefont {Turkeshi}\ \emph {et~al.}(2019)\citenamefont
  {Turkeshi}, \citenamefont {Mendes-Santos}, \citenamefont {Giudici},\ and\
  \citenamefont {Dalmonte}}]{PhysRevLett.122.150606}%
  \BibitemOpen
  \bibfield  {author} {\bibinfo {author} {\bibfnamefont {X.}~\bibnamefont
  {Turkeshi}}, \bibinfo {author} {\bibfnamefont {T.}~\bibnamefont
  {Mendes-Santos}}, \bibinfo {author} {\bibfnamefont {G.}~\bibnamefont
  {Giudici}}, \ and\ \bibinfo {author} {\bibfnamefont {M.}~\bibnamefont
  {Dalmonte}},\ }\href {\doibase 10.1103/PhysRevLett.122.150606} {\bibfield
  {journal} {\bibinfo  {journal} {Phys. Rev. Lett.}\ }\textbf {\bibinfo
  {volume} {122}},\ \bibinfo {pages} {150606} (\bibinfo {year}
  {2019})}\BibitemShut {NoStop}%
\bibitem [{\citenamefont {Zhu}\ \emph {et~al.}(2019)\citenamefont {Zhu},
  \citenamefont {Huang},\ and\ \citenamefont {He}}]{zhu2019reconstructing}%
  \BibitemOpen
  \bibfield  {author} {\bibinfo {author} {\bibfnamefont {W.}~\bibnamefont
  {Zhu}}, \bibinfo {author} {\bibfnamefont {Z.}~\bibnamefont {Huang}}, \ and\
  \bibinfo {author} {\bibfnamefont {Y.-C.}\ \bibnamefont {He}},\ }\href@noop {}
  {\bibfield  {journal} {\bibinfo  {journal} {Phys. Rev. B}\ }\textbf {\bibinfo
  {volume} {99}},\ \bibinfo {pages} {235109} (\bibinfo {year}
  {2019})}\BibitemShut {NoStop}%
\bibitem [{\citenamefont {Cardy}\ and\ \citenamefont
  {Tonni}(2016)}]{Cardy_2016}%
  \BibitemOpen
  \bibfield  {author} {\bibinfo {author} {\bibfnamefont {J.}~\bibnamefont
  {Cardy}}\ and\ \bibinfo {author} {\bibfnamefont {E.}~\bibnamefont {Tonni}},\
  }\href {\doibase 10.1088/1742-5468/2016/12/123103} {\bibfield  {journal}
  {\bibinfo  {journal} {Journal of Statistical Mechanics: Theory and
  Experiment}\ }\textbf {\bibinfo {volume} {2016}},\ \bibinfo {pages} {123103}
  (\bibinfo {year} {2016})}\BibitemShut {NoStop}%
\bibitem [{\citenamefont {Carrasco}\ \emph {et~al.}(2021)\citenamefont
  {Carrasco}, \citenamefont {Elben}, \citenamefont {Kokail}, \citenamefont
  {Kraus},\ and\ \citenamefont {Zoller}}]{carrasco2021theoretical}%
  \BibitemOpen
  \bibfield  {author} {\bibinfo {author} {\bibfnamefont {J.}~\bibnamefont
  {Carrasco}}, \bibinfo {author} {\bibfnamefont {A.}~\bibnamefont {Elben}},
  \bibinfo {author} {\bibfnamefont {C.}~\bibnamefont {Kokail}}, \bibinfo
  {author} {\bibfnamefont {B.}~\bibnamefont {Kraus}}, \ and\ \bibinfo {author}
  {\bibfnamefont {P.}~\bibnamefont {Zoller}},\ }\href {\doibase
  10.1103/prxquantum.2.010102} {\bibfield  {journal} {\bibinfo  {journal} {PRX
  Quantum}\ }\textbf {\bibinfo {volume} {2}},\ \bibinfo {pages} {010102}
  (\bibinfo {year} {2021})}\BibitemShut {NoStop}%
\bibitem [{Note4()}]{Note4}%
  \BibitemOpen
  \bibinfo {note} {For degenerate levels, these will be resonant Rabi
  oscillations}\BibitemShut {NoStop}%
\bibitem [{\citenamefont {Hislop}\ and\ \citenamefont
  {Longo}(1982)}]{hislop1982}%
  \BibitemOpen
  \bibfield  {author} {\bibinfo {author} {\bibfnamefont {P.~D.}\ \bibnamefont
  {Hislop}}\ and\ \bibinfo {author} {\bibfnamefont {R.}~\bibnamefont {Longo}},\
  }\href {https://projecteuclid.org:443/euclid.cmp/1103921046} {\bibfield
  {journal} {\bibinfo  {journal} {Comm. Math. Phys.}\ }\textbf {\bibinfo
  {volume} {84}},\ \bibinfo {pages} {71} (\bibinfo {year} {1982})}\BibitemShut
  {NoStop}%
\end{thebibliography}%

\end{document}